\documentclass[12pt]{iopart}

\usepackage{iopams} 
\expandafter\let\csname equation*\endcsname\relax
\expandafter\let\csname endequation*\endcsname\relax
\usepackage{amsmath}
\usepackage{amssymb}

\usepackage{graphicx}
\usepackage{comment}
\usepackage[caption=false]{subfig}
\usepackage{dcolumn}
\usepackage{bm}
\usepackage[square, sort, numbers]{natbib}
\usepackage{url}
\usepackage{tikz}
\usetikzlibrary{shapes}
\usetikzlibrary{positioning}
\usepackage{qcircuit}
\usepackage{braket}
\usepackage{bbm}
\usepackage{esvect}
\usepackage{ulem}
\usepackage[colorlinks=true,backref=false, linktocpage=true,
citecolor=blue,urlcolor=blue,linkcolor=blue,pdfpagemode=UseOutlines]{hyperref}
\usepackage{todonotes}
\usepackage{multirow}
\usepackage[section]{placeins}
\usepackage{booktabs}
\usepackage{appendix}
\usepackage{makecell}

\begin{document}


\title[Quantum Utility-Scale QEM for Quantum Quench]{Quantum Utility-Scale Error Mitigation for Quantum Quench Dynamics in Heisenberg Spin Chains}

\author{Seokwon Choi$^1$, Talal Ahmed Chowdhury$^{2,3}$, Kwangmin Yu$^{4,*}$}

\address{$^1$Department of Physics, Yonsei University, 50 Yonsei-ro, Seodaemun-gu, Seoul 03722, South Korea.}
\address{$^2$Department of Physics, University of Dhaka, P.O. Box 1000, Dhaka, Bangladesh.}
\address{$^3$Department of Physics and Astronomy, University of Kansas, Lawrence, Kansas 66045, USA.}
\address{$^4$Computational Science Department, Brookhaven National Laboratory, Upton, New York 11973, USA.}

\ead{$^*$kyu@bnl.gov}

\begin{abstract}
We propose a quantum error mitigation method termed self-mitigation, which is comparable with zero-noise extrapolation, to achieve quantum utility on near-term, noisy quantum computers. We investigate the effectiveness of several quantum error mitigation strategies, including self-mitigation, by simulating quantum quench dynamics for Heisenberg spin chains with system sizes up to 104 qubits using IBM quantum processors. In particular, we discuss the limitations of zero-noise extrapolation and the advantages offered by self-mitigation at a large scale.
The self-mitigation method shows stable accuracy with the large systems of 104 qubits with more than 3,000 \texttt{CNOT} gates. Also, we combine the discussed quantum error mitigation methods with practical entanglement entropy measuring methods, and it shows a good agreement with the theoretical estimation. Our study illustrates the usefulness of near-term noisy quantum hardware in examining the quantum quench dynamics of many-body systems at large scales, and lays the groundwork for surpassing classical simulations with quantum methods prior to the development of fault-tolerant quantum computers.
\end{abstract}

\vspace{2pc}
\vspace{2pc}



\section{Introduction}\label{sec:intro}

In recent years, unprecedented advancements in quantum computing have occurred, involving both quantum hardware and quantum algorithms. In fact, preliminary experiments on emerging quantum hardware indicate that these systems can outperform their classical counterparts for certain specialized computational tasks. However, demonstrating a comprehensive, significant, and precisely measurable quantum advantage over classical systems remains a formidable challenge, especially when tackling problems of practical relevance. One of the most challenging obstacles is the quantum noise originating from quantum devices. To overcome this, many quantum error correction (QEC) methods have been developed, and several roadmaps to fault-tolerant quantum computing (FTQC), for which QEC is crucial, have been suggested \cite{google2023suppressing, acharya2024quantum, putterman2025hardware}. 
Even with an optimistic outlook, an early FTQC may be available by 2029. 
A recent study showed that 1457 physical qubits (distance 27 surface code) are required to achieve one logical qubit with $10^{-6}$ logical error rate (LER) \cite{acharya2024quantum}. This shows that reaching FTQC, even the early FTQC era, is quite demanding, despite rapid advances in quantum computing and considering current in-service quantum computers have less than 200 qubits. Therefore, it is essential to maximize the efficiency of today’s noisy quantum computers in the pre-FTQC phase and in the early FTQC systems.

On the other hand, simulating fundamental physics with near-term, noisy quantum computers presents major challenges, including high error rates that impact computational accuracy, limited qubit counts that restrict the complexity of systems that can be simulated, and difficulties in maintaining qubit coherence over time. Nonetheless, advances in error mitigation (QEM) methods~\cite{endo-error-mitigation, Temme-error-mitigation, Li-error-mitigation, Kandala-error-mitigation, Berg-error-mitigation, yu2023simulating, Kim-error-mitigation, kim2023evidence} are steadily improving the ability of noisy quantum devices to carry out detailed and accurate simulations of fundamental physics. These developments have demonstrated the practical utility of noisy quantum computers even before the arrival of fully fault-tolerant quantum systems~\cite{kim2023evidence}. However, simulating large quantum systems and accurately extracting observables for more realistic problems on near-term quantum computers still remains out of reach for now.
In addition to measuring observables, measuring the entanglement entropy is also crucial because it provides deep insight into the quantum correlations present in a system and serves as a powerful diagnostic tool in many areas of quantum physics. In non-equilibrium quantum systems, entanglement entropy growth (e.g., after a quench) reflects how quantum information spreads. Therefore, measuring the entanglement entropy is key to understanding thermalization, many-body localization (MBL), and scrambling of quantum information. However, measuring the entanglement entropy on near-term quantum computers is still challenging. Addressing the challenges of measuring observables and entanglement entropy requires further exploration to better understand the practical capabilities and limitations of current quantum hardware for large-scale quantum simulations. 

Therefore, it is essential to efficiently apply QEM methods to the quantum algorithms, as well as develop more efficient QEM techniques, to enable the effective use of near-term noisy quantum computers at the utility scale, before the advent of fully FTQC, even in its early stages.
Hence, we advance a quantum error mitigation method termed self-mitigation \cite{rahman2022self}, which is comparable with zero-noise extrapolation \cite{Temme-error-mitigation, Li-error-mitigation, giurgica2020digital} to achieve quantum utility on near-term, noisy quantum computers.
We extend the self-mitigation method to the optimized second-order Trotterization \cite{chowdhury2024enhancing} for the XXZ spin chain model.
We evaluate several quantum error mitigation techniques, including self-mitigation, and analyze their respective strengths and weaknesses. In particular, we analyze the limitations of zero-noise extrapolation and the advantages offered by self-mitigation at a large scale. Our results demonstrate that self-mitigation surpasses zero-noise extrapolation in both accuracy and resource efficiency. Building on this benchmark analysis ($20$-qubit system), we extend our experiments to larger systems, scaling up to $84$ qubits with periodic boundary conditions and $104$ qubits with open boundary conditions. The extended simulations show a good accuracy with more than 3,000 \texttt{CNOT} gates. The self-mitigation method shows stable accuracy with the large systems, while the zero-noise extrapolation method has a higher error rate than the benchmark test of $20$-qubits. 
Finally, we explore computing the entanglement entropy after time evolution of the Hamiltonian. We address a practical protocol for the entropy measurement on quantum devices and implement it on IBM quantum computers. The entropy measurement result shows a good agreement with the classical result.

We focus on studying the quench dynamics in the one-dimensional anisotropic spin-$\frac{1}{2}$ antiferromagnetic Heisenberg model or XXZ spin chain using IBM superconducting quantum computers and assess their ability to accurately capture the intricate spin dynamics of the model. The XXZ model is a paradigmatic example of a quantum spin system that exhibits integrability and critical behavior~\cite{Orbach, Lieb-Schultz-Mattis, Gaudin, Yang-Yang} where the critical regime of the XXZ spin chain can be described by a $1+1$D conformal field theory with central charge $c=1$~\cite{Alcaraz, DEVEGA1985439}. Besides, the low-energy limit of the XXZ spin chain can be connected to lower-dimensional quantum field theories, as shown in Refs.~\cite{Luther-Peschel, LUKYANOV1998533}. In addition, the time evolution in the XXZ spin chain after the quantum quenches at zero temperature is studied in Refs.~\cite{Calabrese-Cardy-1, DeChiara, Fagotti} which focused on dynamics of Von Neumann entropy of a block of spins and spin-spin correlation functions after a sudden quench in the anisotropy parameter, and for finite temperature case in Ref.~\cite{Bonnes} which studied the spreading of information after quenches from a variety of initial thermal density matrices. For a detailed discussion on quench dynamics and relaxation in isolated integrable quantum spin chains, refer to Ref.~\cite{Essler-review}.

The following section reviews the quantum circuit implementation of the XXZ spin chain Hamiltonian used in this study. Section \ref{sec:error_mitigations} presents an overview of the quantum error mitigation techniques used in this study, particularly emphasizing our improved self-mitigation technique in section \ref{sec:SM}. In \sref{sec:observable_measurement}, we compare these methods and evaluate their effectiveness. Finally, we evaluate the entanglement entropy of the quantum state after its time evolution governed by the Hamiltonian.

\section{Implementation of Time Evolution for the XXZ spin chain Hamiltonian}\label{sec:time-evoliton}

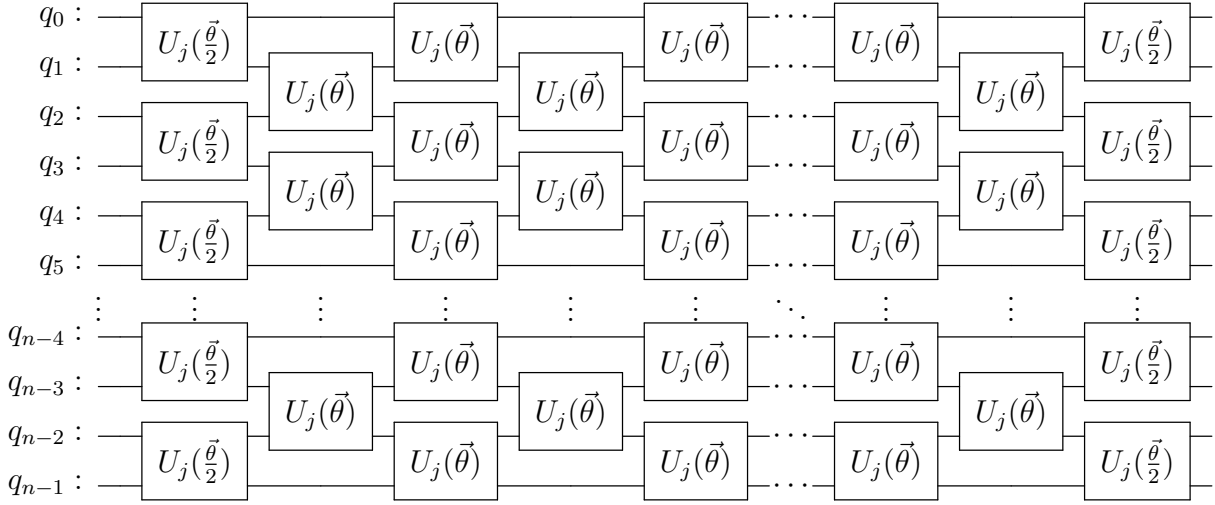
\begin{figure}[t!]
\[
\Qcircuit @C=0.7em @R=0.7em {
\lstick{ {q}_{0} :  }   & \qw & \multigate{1}{U_j (\frac{\vec{\theta}}{2})} & \qw & \multigate{1}{U_j (\vec{\theta})} & \qw & \multigate{1}{U_j (\vec{\theta})} & \qw & \cdots & & \multigate{1}{U_j (\vec{\theta})} & \qw & \multigate{1}{U_j (\frac{\vec{\theta}}{2})} & \qw \\
\lstick{ {q}_{1} :  }   & \qw & \ghost{U_j (\frac{\vec{\theta}}{2})} & \multigate{1}{U_j (\vec{\theta})} & \ghost{U_j (\vec{\theta})} & \multigate{1}{U_j (\vec{\theta})} & \ghost{U_j (\vec{\theta})} & \qw & \cdots & & \ghost{U_j (\vec{\theta})} & \multigate{1}{U_j (\vec{\theta})} & \ghost{U_j (\frac{\vec{\theta}}{2})} & \qw \\
\lstick{ {q}_{2} :  }   & \qw & \multigate{1}{U_j (\frac{\vec{\theta}}{2})} & \ghost{U_j (\vec{\theta})} &  \multigate{1}{U_j (\vec{\theta})} & \ghost{U_j (\vec{\theta})}  & \multigate{1}{U_j (\vec{\theta})} & \qw & \cdots & &  \multigate{1}{U_j (\vec{\theta})} & \ghost{U_j (\vec{\theta})}  & \multigate{1}{U_j (\frac{\vec{\theta}}{2})} & \qw \\
\lstick{ {q}_{3} :  }   & \qw & \ghost{U_j (\frac{\vec{\theta}}{2})} & \multigate{1}{U_j (\vec{\theta})} &  \ghost{U_j (\vec{\theta})} & \multigate{1}{U_j (\vec{\theta})} & \ghost{U_j (\vec{\theta})} & \qw & \cdots & &  \ghost{U_j (\vec{\theta})} & \multigate{1}{U_j (\vec{\theta})} & \ghost{U_j (\frac{\vec{\theta}}{2})} & \qw \\
\lstick{ {q}_{4} :  }   & \qw & \multigate{1}{U_j (\frac{\vec{\theta}}{2})} & \ghost{U_j (\vec{\theta})} &  \multigate{1}{U_j (\vec{\theta})} & \ghost{U_j (\vec{\theta})} & \multigate{1}{U_j (\vec{\theta})} & \qw & \cdots & &  \multigate{1}{U_j (\vec{\theta})} & \ghost{U_j (\vec{\theta})} & \multigate{1}{U_j (\frac{\vec{\theta}}{2})} & \qw  \\
\lstick{ {q}_{5} :  }   & \qw & \ghost{U_j (\frac{\vec{\theta}}{2})} & \qw & \ghost{U_j (\vec{\theta})} & \qw & \ghost{U_j (\vec{\theta})}& \qw & \cdots & & \ghost{U_j (\vec{\theta})} & \qw & \ghost{U_j (\frac{\vec{\theta}}{2})}& \qw \\
\vdots  &  & \vdots & \vdots & \vdots & \vdots & \vdots &  & \ddots & & \vdots & \vdots & \vdots &  \\
\lstick{ {q}_{n-4} :  } & \qw & \multigate{1}{U_j (\frac{\vec{\theta}}{2})} & \qw & \multigate{1}{U_j (\vec{\theta})} & \qw & \multigate{1}{U_j (\vec{\theta})} & \qw & \cdots & & \multigate{1}{U_j (\vec{\theta})} & \qw & \multigate{1}{U_j (\frac{\vec{\theta}}{2})} & \qw \\
\lstick{ {q}_{n-3} :  } & \qw & \ghost{U_j (\frac{\vec{\theta}}{2})}& \multigate{1}{U_j (\vec{\theta})} & \ghost{U_j (\vec{\theta})} & \multigate{1}{U_j (\vec{\theta})} & \ghost{U_j (\vec{\theta})}& \qw & \cdots & & \ghost{U_j (\vec{\theta})} & \multigate{1}{U_j (\vec{\theta})} & \ghost{U_j (\frac{\vec{\theta}}{2})}& \qw \\
\lstick{ {q}_{n-2} :  } & \qw & \multigate{1}{U_j (\frac{\vec{\theta}}{2})} & \ghost{U_j (\vec{\theta})} & \multigate{1}{U_j (\vec{\theta})} & \ghost{U_j (\vec{\theta})} & \multigate{1}{U_j (\vec{\theta})} & \qw  & \cdots & & \multigate{1}{U_j (\vec{\theta})} & \ghost{U_j (\vec{\theta})} & \multigate{1}{U_j (\frac{\vec{\theta}}{2})} & \qw \\
\lstick{ {q}_{n-1} :  } & \qw & \ghost{U_j (\frac{\vec{\theta}}{2})} & \qw & \ghost{U_j (\vec{\theta})} & \qw & \ghost{U_j (\vec{\theta})}& \qw & \cdots & & \ghost{U_j (\vec{\theta})} & \qw & \ghost{U_j (\frac{\vec{\theta}}{2})}& \qw
}
\]
\vspace*{1mm}
\caption{The optimized second-order Trotterization of the Hamiltonian for the XXZ spin chain model with open boundary conditions. The index $j$ represents the first qubit index where $U_j$ is placed. For a periodic boundary condition, the odd layers have the two-qubit gates, $U_j (\vec{\theta})$, between $q_{n-1}$ and $q_0$.}
\label{fig:circuit_XXX_2nd_opt}
\end{figure}

The Heisenberg XXZ spin chain model is described as follows:
\begin{equation}
    H = J_1 \sum_{i=1}^{N}\left(S^{x}_{i}S^{x}_{i+1}+S^{y}_{i}S^{y}_{i+1}+\Delta S^{z}_{i}S^{z}_{i+1}\right)
\label{eq:hamiltonian}
\end{equation}
where the antiferromagnetic nearest-neighbor coupling $J_1 > 0$ and the exchange-anisotropy parameter $\Delta$ controls the parameter space of the Hamiltonian. The system exhibits a ferromagnetic phase when $\Delta <-1$, a critical phase when $-1<\Delta<1$ and an antiferromagnetic phase when $\Delta >1$. Besides, the spin operators, $S^{i}=\frac{1}{2}\sigma^{i}$ obey the $SU(2)$ algebra,
\begin{equation}
\notag
[S^{\alpha}_{i},S^{\beta}_{j}]=i\delta_{ij}\epsilon^{\alpha\beta\gamma}S^{\gamma}_{i},
\end{equation}
where $\alpha,\,\beta,\,\gamma=x,\,y,\,z$ and $i,\,j=1,...,N$.
\Eref{eq:hamiltonian} is reformulated by the Pauli operators, $\sigma^x_j,\sigma^y_j,\sigma^z_j$ as follows:
\begin{equation}\label{eq:hamiltonian_Pauli}
H = \frac{J_1}{4} \sum_{j=0}^{N-1} 
 \Bigl(\sigma_j^x\sigma_{j+1}^x + \sigma_j^y\sigma_{j+1}^y +  \Delta \sigma_j^z\sigma_{j+1}^z \Bigr).
\end{equation}
To execute the time evolution driven by the Hamiltonian in equation \eref{eq:hamiltonian_Pauli}, we use the second-order Trotterization~\cite{trotter, suzuki1, suzuki2} as follows:
\begin{equation}
    U_{tot}(t)=e^{-i H t}\approx \left( \prod_{n=1}^{M}e^{-i H_{n}t/{2r}} \prod_{n=M}^{1}e^{-i H_{n}t/{2r}} \right)^{r},
    \label{eq:second-trotterization}
\end{equation}
where $H = \sum H_{n}$ and $r$ is the number of Trotter steps. The Trotter timestep size is defined as $\delta  t = t/r$. 
We adopt an efficient second-order Trotterization for the model that is described in Ref. \cite{chowdhury2024enhancing}.
The circuit diagram is visualized in \fref{fig:circuit_XXX_2nd_opt}.

The second-order Trotterization is composed of the following basic building blocks.
\begin{equation}
\label{eq:evolution_XYZ}
U_j (\vec{\theta}) = \exp \left(  -i \left( \frac{\theta_x}{2} \sigma_j^x \sigma_{j+1}^x + \frac{\theta_y}{2} \sigma_j^y \sigma_{j+1}^y + \frac{\theta_z}{2} \sigma_j^z \sigma_{j+1}^z \right) \right) ,
\end{equation}
where $\vec{\theta} = (\theta_x, \theta_y, \theta_z) = (\frac{J_1}{2} \delta t, ~ \frac{J_1}{2} \delta t, ~ \frac{J_1}{2} \Delta \delta t)$ with the Trotter step size $\delta t$ acting on qubit $j$ and $j+1$.
By the Trotter approximation, we approximate the time evolution $U_{tot}(t)$ in equation \eref{eq:second-trotterization} by arranging the $U_j (\vec{\theta})$ operators in staggered placement (brick-wall structure) \cite{vanicat2018integrable, smith2019simulating} and finally optimized as shown in \fref{fig:circuit_XXX_2nd_opt}.
Note that the optimized second-order Trotterization has only $2M + 1$ layers of $U_j$ building blocks with $M$ Trotter steps, while the first-order Trotterization has $2M$ layers with $M$ Trotter steps.
Hence, the optimized second-order Trotterization is implemented by adding one even layer at the end of the first-order Trotterization and adjusting the angle parameters ($\vec{\theta}$).

\subsection{Unit Block Implementation}\label{sec:quantum_circuit}

Since $U_j (\vec{\theta})$ in equation~\eref{eq:evolution_XYZ} is the basic building block for the quantum circuit implementation as discussed in the previous section, achieving reliable simulations on noisy quantum computers critically depends on the design and implementation of efficient quantum circuits, given that near-term noisy devices are highly susceptible to multiple sources of quantum noise, particularly gate errors.

Based on \texttt{RZZGate} in IBM Qiskit, we construct the induced $R_{X_i X_j} (\theta)$ and $R_{Y_i Y_j} (\theta)$ gates by the Clifford gate identities as follows:

\begin{align}
\notag
    \Qcircuit @C=0.7em @R=1.5em {
    &&&&& \gate{H} & \ctrl{1} & \qw & \ctrl{1} & \gate{H} & \\
    \raisebox{1.2cm}{$R_{XX}(\theta)=$} &&&&& \gate{H} & \targ & \gate{R_z(\theta)} & \targ & \gate{H} & , \\
    &&&&& \gate{\sqrt{\sigma^x}} & \ctrl{1} & \qw & \ctrl{1} & \gate{\sqrt{\sigma^x}^{\dagger}} & \\
    \raisebox{1.2cm}{$R_{YY}(\theta)=$} &&&&& \gate{\sqrt{\sigma^x}} & \targ & \gate{R_z(\theta)} & \targ & \gate{\sqrt{\sigma^x}^{\dagger}} & , \\
    &&&&&& \ctrl{1} & \qw & \ctrl{1} & \qw & \\
    \raisebox{0.8cm}{$R_{ZZ}(\theta)=$} &&&&&& \targ & \gate{R_z(\theta)} & \targ & \qw & 
    }
\end{align}
where $H$ is the Hadarmard gate, $\sqrt{\sigma^x} = \frac{1}{2} {\begin{pmatrix} 1+i & 1-i \\ 1-i & 1+i \end{pmatrix}}$, and $R_z (\theta) = {\begin{pmatrix} e^{-i \frac{\theta}{2}} & 0 \\ 0 & e^{i \frac{\theta}{2}} \end{pmatrix}}$.
Hence, $U_i (\vec{\theta})$ in Eq.~\eqref{eq:evolution_XYZ} is implemented as follows:
\begin{equation}
\notag
    \Qcircuit @C=0.7em @R=1.5em {
    &&& \gate{H} & \ctrl{1} & \qw & \ctrl{1} & \gate{H} & \qw & \gate{\sqrt{\sigma^x}} & \qw & \ctrl{1} & \qw & \ctrl{1} & \gate{\sqrt{\sigma^x}^{\dagger}} & \ctrl{1} & \qw & \ctrl{1} & \qw & \\
    &&& \gate{H} & \targ & \gate{R_z(\theta)} & \targ & \gate{H} & \qw & \gate{\sqrt{\sigma^x}} & \qw & \targ & \gate{R_z(\theta)} & \targ & \gate{\sqrt{\sigma^x}^{\dagger}} & \targ & \gate{R_z(\theta)} & \targ & \qw & 
    }
\end{equation}
and this implementation has six \texttt{CX} gates and thirteen circuit depths.
This circuit is compressed and optimized by circuit identities as follows:

\begin{equation}
\label{eq:circuit_XYZ}
\Qcircuit @C=1.0em @R=0.7em @!R { 
 & \targ & \gate{R_z ( \theta_z )} & \qw & \targ & \gate{R_z (- \theta_y)} & \targ & \gate{\sqrt{\sigma^x}} & \qw & \qw & \\
 & \ctrl{-1} & \gate{H} & \gate{R_z (\theta_x + \frac{\pi}{2})} & \ctrl{-1} & \gate{H} & \ctrl{-1} &  \gate{\sqrt{\sigma^x}^\dagger} & \qw & \qw  
}
\end{equation}
and this circuit has three \texttt{CX} gates and seven depths.
The derivation of the circuit identity is detailed in Appendix A of Ref. \cite{zhang2024optimal}.
Given that quantum gates are prone to errors, particularly two-qubit gates like \texttt{CX}, which tend to be significantly noisier than single-qubit gates. minimizing both the number of \texttt{CX} gates and the overall circuit depth is crucial for mitigating noise in quantum computations.
Therefore, we utilize the quantum circuit defined in equation~\eref{eq:circuit_XYZ} for the implementation of Eq.~\eqref{eq:evolution_XYZ}.

\section{Quantum Error Mitigations}\label{sec:error_mitigations}

A major challenge in running quantum algorithms on contemporary quantum devices, such as IBM Quantum processors, is the presence of errors and noise. To address these issues, researchers have proposed quantum error correction (QEC) techniques. 
However, QEC requires a significant qubit overhead, making it impractical to implement for large-scale problems on near-term noisy quantum processors, even with optimizations \cite{kivlichan2020improved, lee2021even}.
In contrast, quantum error mitigation (QEM) embraces the imperfections of current quantum devices and employs techniques to reduce or suppress quantum errors and noise. Unlike QEC, QEM typically incurs little to no qubit overhead. In recent years, a variety of QEM methods have been developed and demonstrated to be effective in solving practical problems \cite{yu2023simulating, Kim-error-mitigation, kim2023evidence, charles2305simulating, chowdhury2024enhancing}. 
To mitigate quantum device errors and noise in our experiments, we employ five QEM techniques: Twirled Readout Error Extinction (TREX), Dynamical Decoupling (DD), Pauli Twirling (PT), Zero-Noise Extrapolation (ZNE), and Self-Mitigation (SM). The details of each QEM method are provided in the following sub-sections.

\subsection{Twirled Readout Error Extinction} 
Twirled Readout Error Extinction (TREX) is a model-free technique for mitigating readout errors in quantum circuits \cite{PhysRevA.105.032620}. It randomly applies Pauli-X gates to qubits just before measurement and then flips the corresponding classical bits in the output. In an ideal, noise-free setting, this process has no effect—it is effectively the identity. However, under realistic noise conditions, it symmetrizes the readout error channel, transforming complex, state-dependent bias into a simpler, uniform scaling factor.

TREX transforms arbitrary readout errors into a diagonal form by randomizing the measurement basis. This makes the error model much simpler (each qubit's error is independent and only affects its own measurement outcome), allowing for straightforward inversion and correction.
Also, TREX does not require detailed modeling of the underlying noise, making it robust even when the noise is hard to characterize.
However, it is less effective when the measurement errors are strongly correlated across qubits.
In our experiments, we adopt $10$ sampling numbers for the TREX.

\subsection{Dynamical Decoupling}

Dynamical Decoupling (DD) is a quantum error mitigation technique designed to suppress errors arising from interactions with spectator qubits.
It involves applying periodic sequences of instantaneous control pulses that effectively average out the system’s coupling to the environment, driving it toward zero \cite{viola1999dynamical}.
Specifically, DD interleaves single-qubit operations on idle qubits through basis transformations, thereby isolating them from environmental noise introduced by other qubits.
Consequently, the coherence time of the circuit becomes longer.
The efficiency of DD is empirically tested in various environments \cite{ezzell2022dynamical, niu2022effects, Kim-error-mitigation, kim2023evidence, charles2305simulating}.
In this study, we use ($t/4$, $X$, $t/2$, $X$, $t/4$) sequence in every idling period for the DD implementation, where $X$ represents the $\texttt{XGate}$ and $t$ is the idling time except for the two $\texttt{XGate}$ pulse durations.

\subsection{Pauli Twirling}

Pauli Twirling (PT) is a technique used to transform coherent errors in quantum circuits into stochastic Pauli errors by averaging out the off-diagonal components in the Pauli basis, $\{ I, \sigma^x, \sigma^y, \sigma^z \}$ \cite{bennett1996purification, wallman2016noise, cai2019constructing}.
In PT, a Clifford gate is surrounded by the Pauli gates back and forth, which is mathematically identical to the Clifford gate.
The efficiency is empirically proved in previous studies \cite{Kim-error-mitigation, kim2023evidence, chowdhury2024enhancing, chowdhury2024capturing, chowdhury2025first}.
We apply the PT only to two-qubit Clifford gates, $\texttt{CX}$ and $\texttt{CZ}$ gates.
We duplicated $10$ copies of the base quantum circuit.
We randomly chose a Pauli twirling gate combination from the prepared Pauli twirling gates set and applied them to the two-qubit Clifford gates.

\subsection{Zero-Noise Extrapolation}

Zero-Noise Extrapolation (ZNE) mitigates quantum errors by inferring the noise-free expectation value through extrapolation from a set of noisy measurements collected at different noise strengths \cite{Temme-error-mitigation, Li-error-mitigation, giurgica2020digital}.
For our experiments, local unitary gate folding \cite{giurgica2020digital} is selectively applied to the two-qubit gates, $\texttt{CX}$, using scaling factors of 1, 3, and 5, given that the two-qubit gates are more than an order of magnitude noisier than single-qubit gates.

ZNE relies on the assumption that noise levels can be systematically increased in a controlled manner, despite the potential nonlinearity of the extrapolation curve.
However, in reality, quantum noise is often non-Markovian and time-varying, meaning that scaling the noise doesn't always produce predictable changes in the output. If the noise model is inaccurate or unstable, the extrapolation can become unreliable or even misleading.
When the exact value is known, we can construct an accurate fitting curve for extrapolation. This curve, derived from small-scale estimations, can then be extended to larger scales under the assumption that the noise model remains consistent during scaling. However, this assumption often breaks down when the number of qubits changes. Therefore, we explore an alternative quantum error mitigation (QEM) method in the following section.

\subsection{Self-Mitigation}
\label{sec:SM}

The self-mitigating (SM) QEM method, suggested in Ref. \cite{rahman2022self}, represents a specific error mitigation strategy particularly tailored for Trotterization. The self-mitigation is a special case of the depolarizing noise mitigation method suggested in Ref. \cite{urbanek2021mitigating}.
The depolarizing noise mitigation method requires a test circuit to measure the noise probability (or noise factor) $p$.
The relation between the true expectation value $\left<O\right>$ of an observable $O$ and a noisy expectation value $\overline{\left<O\right>}$ of the observable $O$ is described as follows \cite{urbanek2021mitigating}:
\begin{align}
\notag
\left<O\right> = \frac{\overline{\left<O\right>}}{1 - p}.
\end{align}
where $p$ is the noise factor.
Particularly, the self-mitigation uses the time discretization alternatively in the Trotter step with $dt$ and $-dt$ for the test quantum circuit measuring $\overline{\left<O\right>}$.
For example, the test circuit has $(dt, -dt, ~dt, -dt, \cdots, ~dt, -dt)$ time sequence while the target circuit (the original circuit of concern) has $(dt, ~dt, ~dt, ~dt, \cdots, ~dt, ~dt)$ at the Trotter steps.
Since the test circuit applies $dt$ and $-dt$ alternatively in the Trotter steps, the ideal (noiseless) quantum state after the Trotter steps is the initial state, and it is already known or relatively easier to estimate the expectation value of the observable at the initial state.
Hence, we have $\left<O\right>$ by the (ideal) initial state estimation and $\overline{\left<O\right>}$ by running the test circuit.
Then we estimate the noise factor $p$.
Lastly, we run the target Trotterization circuit (forwarding with $dt$) and obtain the noisy expectation value $\overline{\left<O\right>}$ of the target circuit. After applying the noise factor $p$, we estimate the noiseless expectation value $\left<O\right>$.
Note that the discrepancy between the noise factors of the test circuit and the target circuit is negligible to justify the depolarizing noise mitigation method.
Since the self-mitigation generates a test circuit with $-dt$ instead of $dt$, the test circuit preserves the circuit structure of the target circuit. The test circuit has the reversed angle parameter by $-dt$ instead of $dt$. Hence, we assume the discrepancy between the noise factors of the test circuit and the target circuit is negligible.

\begin{figure}[t!]
\[
\Qcircuit @C=0.4em @R=0.7em {
\lstick{ {q}_{0} :  }   & \qw & \multigate{1}{U_j (\frac{\vec{\theta}}{2})} & \qw & \multigate{1}{U_j (\vec{\textbf{0}})} & \qw & \multigate{1}{U_j (\vec{\textbf{0}})} & \qw & \multigate{1}{U_j (\vec{\textbf{0}})} & \qw & \multigate{1}{U_j (-\frac{\vec{\theta}}{2})} & \qw\\
\lstick{ {q}_{1} :  }   & \qw & \ghost{U_j (\frac{\vec{\theta}}{2})} & \multigate{1}{U_j (\vec{\theta})} & \ghost{U_j (\vec{\textbf{0}})} & \multigate{1}{U_j (-\vec{\theta})} & \ghost{U_j (\vec{\textbf{0}})} & \multigate{1}{U_j (\vec{\theta})} & \ghost{U_j (\vec{\textbf{0}})} & \multigate{1}{U_j (-\vec{\theta})} & \ghost{U_j (-\frac{\vec{\theta}}{2})} & \qw\\
\lstick{ {q}_{2} :  }   & \qw & \multigate{1}{U_j (\frac{\vec{\theta}}{2})} & \ghost{U_j (\vec{\theta})} &  \multigate{1}{U_j (\vec{\textbf{0}})} & \ghost{U_j (-\vec{\theta})}  & \multigate{1}{U_j (\vec{\textbf{0}})} & \ghost{U_j (\vec{\theta})}  & \multigate{1}{U_j (\vec{\textbf{0}})} &  \ghost{U_j (-\vec{\theta})}  & \multigate{1}{U_j (-\frac{\vec{\theta}}{2})} & \qw \\
\lstick{ {q}_{3} :  }   & \qw & \ghost{U_j (\frac{\vec{\theta}}{2})} & \multigate{1}{U_j (\vec{\theta})} &  \ghost{U_j (\vec{\textbf{0}})} & \multigate{1}{U_j (-\vec{\theta})} & \ghost{U_j (\vec{\textbf{0}})} &  \multigate{1}{U_j (\vec{\theta})} & \ghost{U_j (\vec{\textbf{0}})} & \multigate{1}{U_j (-\vec{\theta})} & \ghost{U_j (-\frac{\vec{\theta}}{2})} & \qw \\
\lstick{ {q}_{4} :  }   & \qw & \multigate{1}{U_j (\frac{\vec{\theta}}{2})} & \ghost{U_j (\vec{\theta})} &  \multigate{1}{U_j (\vec{\textbf{0}})} & \ghost{U_j (-\vec{\theta})} & \multigate{1}{U_j (\vec{\textbf{0}})} &  \ghost{U_j (\vec{\theta})} & \multigate{1}{U_j (\vec{\textbf{0}})} & \ghost{U_j (-\vec{\theta})} & \multigate{1}{U_j (-\frac{\vec{\theta}}{2})} & \qw \\
\lstick{ {q}_{5} :  }   & \qw & \ghost{U_j (\frac{\vec{\theta}}{2})} & \qw & \ghost{U_j (\vec{\textbf{0}})} & \qw & \ghost{U_j (\vec{\textbf{0}})}& \qw &  \ghost{U_j (\vec{\textbf{0}})}& \qw & \ghost{U_j (-\frac{\vec{\theta}}{2})}& \qw \\
\vdots  &  & \vdots & \vdots & \vdots & \vdots & \vdots &  \vdots & \vdots & \vdots & \vdots & \\
\lstick{ {q}_{n-4} :  } & \qw & \multigate{1}{U_j (\frac{\vec{\theta}}{2})} & \qw & \multigate{1}{U_j (\vec{\textbf{0}})} & \qw & \multigate{1}{U_j (\vec{\textbf{0}})} & \qw & \multigate{1}{U_j (\vec{\textbf{0}})} & \qw & \multigate{1}{U_j (-\frac{\vec{\theta}}{2})} & \qw \\
\lstick{ {q}_{n-3} :  } & \qw & \ghost{U_j (\frac{\vec{\theta}}{2})}& \multigate{1}{U_j (\vec{\theta})} & \ghost{U_j (\vec{\textbf{0}})} & \multigate{1}{U_j (-\vec{\theta})} & \ghost{U_j (\vec{\textbf{0}})}& \multigate{1}{U_j (\vec{\theta})} & \ghost{U_j (\vec{\textbf{0}})}& \multigate{1}{U_j (-\vec{\theta})} & \ghost{U_j (-\frac{\vec{\theta}}{2})}& \qw \\
\lstick{ {q}_{n-2} :  } & \qw & \multigate{1}{U_j (\frac{\vec{\theta}}{2})} & \ghost{U_j (\vec{\theta})} & \multigate{1}{U_j (\vec{\textbf{0}})} & \ghost{U_j (-\vec{\theta})} & \multigate{1}{U_j (\vec{\textbf{0}})} & \ghost{U_j (\vec{\theta})} & \multigate{1}{U_j (\vec{\textbf{0}})} & \ghost{U_j (-\vec{\theta})} & \multigate{1}{U_j (-\frac{\vec{\theta}}{2})} & \qw \\
\lstick{ {q}_{n-1} :  } & \qw & \ghost{U_j (\frac{\vec{\theta}}{2})} & \qw & \ghost{U_j (\vec{\textbf{0}})} & \qw & \ghost{U_j (\vec{\textbf{0}})}& \qw & \ghost{U_j (\vec{\textbf{0}})}& \qw & \ghost{U_j (-\frac{\vec{\textbf{0}}}{2})}& \qw
}
\]
\vspace*{1mm}
\caption{The test circuit diagram of $(dt, -dt, ~dt, -dt)$ time sequence. This test circuit has $(dt, -dt, ~dt, -dt)$ time sequence for the optimized second-order Trotterization (\fref{fig:circuit_XXX_2nd_opt}).}
\label{fig:self_mitigation1}
\end{figure}

\begin{figure}[t!]
\[
\Qcircuit @C=0.4em @R=0.7em {
\lstick{ {q}_{0} :  }   & \qw & \multigate{1}{U_j (\frac{\vec{\theta}}{2})} & \qw & \multigate{1}{U_j (\vec{\theta})} & \qw & \multigate{1}{U_j (\vec{\textbf{0}})} & \qw & \multigate{1}{U_j (-\vec{\theta})} & \qw & \multigate{1}{U_j (-\frac{\vec{\theta}}{2})} & \qw\\
\lstick{ {q}_{1} :  }   & \qw & \ghost{U_j (\frac{\vec{\theta}}{2})} & \multigate{1}{U_j (\vec{\theta})} & \ghost{U_j (\vec{\theta})} & \multigate{1}{U_j (\vec{\theta})} & \ghost{U_j (\vec{\textbf{0}})} & \multigate{1}{U_j (-\vec{\theta})} & \ghost{U_j (-\vec{\theta})} & \multigate{1}{U_j (-\vec{\theta})} & \ghost{U_j (-\frac{\vec{\theta}}{2})} & \qw\\
\lstick{ {q}_{2} :  }   & \qw & \multigate{1}{U_j (\frac{\vec{\theta}}{2})} & \ghost{U_j (\vec{\theta})} &  \multigate{1}{U_j (\vec{\theta})} & \ghost{U_j (\vec{\theta})}  & \multigate{1}{U_j (\vec{\textbf{0}})} & \ghost{U_j (-\vec{\theta})}  & \multigate{1}{U_j (-\vec{\theta})} &  \ghost{U_j (-\vec{\theta})}  & \multigate{1}{U_j (-\frac{\vec{\theta}}{2})} & \qw \\
\lstick{ {q}_{3} :  }   & \qw & \ghost{U_j (\frac{\vec{\theta}}{2})} & \multigate{1}{U_j (\vec{\theta})} &  \ghost{U_j (\vec{\theta})} & \multigate{1}{U_j (\vec{\theta})} & \ghost{U_j (\vec{\textbf{0}})} &  \multigate{1}{U_j (-\vec{\theta})} & \ghost{U_j (-\vec{\theta})} & \multigate{1}{U_j (-\vec{\theta})} & \ghost{U_j (-\frac{\vec{\theta}}{2})} & \qw \\
\lstick{ {q}_{4} :  }   & \qw & \multigate{1}{U_j (\frac{\vec{\theta}}{2})} & \ghost{U_j (\vec{\theta})} &  \multigate{1}{U_j (\vec{\theta})} & \ghost{U_j (\vec{\theta})} & \multigate{1}{U_j (\vec{\textbf{0}})} &  \ghost{U_j (-\vec{\theta})} & \multigate{1}{U_j (-\vec{\theta})} & \ghost{U_j (-\vec{\theta})} & \multigate{1}{U_j (-\frac{\vec{\theta}}{2})} & \qw \\
\lstick{ {q}_{5} :  }   & \qw & \ghost{U_j (\frac{\vec{\theta}}{2})} & \qw & \ghost{U_j (\vec{\theta})} & \qw & \ghost{U_j (\vec{\textbf{0}})}& \qw &  \ghost{U_j (\frac{-\vec{\theta}}{2})}& \qw & \ghost{U_j (-\frac{\vec{\theta}}{2})}& \qw \\
\vdots  &  & \vdots & \vdots & \vdots & \vdots & \vdots &  \vdots & \vdots & \vdots & \vdots & \\
\lstick{ {q}_{n-4} :  } & \qw & \multigate{1}{U_j (\frac{\vec{\theta}}{2})} & \qw & \multigate{1}{U_j (\vec{\theta})} & \qw & \multigate{1}{U_j (\vec{\textbf{0}})} & \qw & \multigate{1}{U_j (-\vec{\theta})} & \qw & \multigate{1}{U_j (-\frac{\vec{\theta}}{2})} & \qw \\
\lstick{ {q}_{n-3} :  } & \qw & \ghost{U_j (\frac{\vec{\theta}}{2})}& \multigate{1}{U_j (\vec{\theta})} & \ghost{U_j (\vec{\theta})} & \multigate{1}{U_j (\vec{\theta})} & \ghost{U_j (\vec{\textbf{0}})}& \multigate{1}{U_j (-\vec{\theta})} & \ghost{U_j (-\vec{\theta})}& \multigate{1}{U_j (-\vec{\theta})} & \ghost{U_j (-\frac{\vec{\theta}}{2})}& \qw \\
\lstick{ {q}_{n-2} :  } & \qw & \multigate{1}{U_j (\frac{\vec{\theta}}{2})} & \ghost{U_j (\vec{\theta})} & \multigate{1}{U_j (\vec{\theta})} & \ghost{U_j (\vec{\theta})} & \multigate{1}{U_j (\vec{\textbf{0}})} & \ghost{U_j (-\vec{\theta})} & \multigate{1}{U_j (-\vec{\theta})} & \ghost{U_j (-\vec{\theta})} & \multigate{1}{U_j (-\frac{\vec{\theta}}{2})} & \qw \\
\lstick{ {q}_{n-1} :  } & \qw & \ghost{U_j (\frac{\vec{\theta}}{2})} & \qw & \ghost{U_j (\vec{\theta})} & \qw & \ghost{U_j (\vec{\textbf{0}})}& \qw & \ghost{U_j (-\vec{\theta})}& \qw & \ghost{U_j (-\frac{\vec{\theta}}{2})}& \qw
}
\]
\vspace*{1mm}
\caption{The test circuit diagram of $(dt, ~dt, -dt, -dt)$ time sequence. This test circuit has $(dt, ~dt, -dt, -dt)$ time sequence for the optimized second-order Trotterization (\fref{fig:circuit_XXX_2nd_opt}).}
\label{fig:self_mitigation2}
\end{figure}

We implemented the test circuit of the SM on the optimized second-order Trotterization (refer to \fref{fig:circuit_XXX_2nd_opt}).
When we apply $(dt, -dt, ~dt, -dt, \cdots, ~dt, -dt)$ for the test circuit, the circuit parameters $\vec{\theta}$ in the $U_j(\vec{\theta})$ of even layers become $\vec{0}$ except for the first and the last layer as shown in  \fref{fig:self_mitigation1}, which has $(dt, ~dt, -dt, -dt)$ time sequence.
Since $U_j(\vec{0})$ is the identity operation, all the intermediate operations ($U_j(\vec{0})$) will vanish. In this case, due to the considerable structural mismatch between the test and target circuits, the error rate, $p$, measured from the test circuit does not accurately reflect that of the target circuit.
Therefore, we adopt $(dt, ~dt, ~dt, ~dt, \cdots, -dt, -dt, -dt, -dt)$ time sequence for the test circuit.
In other words, we apply the $dt$ for the first half sequence and the $-dt$ for the last half sequence.
\Fref{fig:self_mitigation2} visualizes the test circuit implementation having $(dt, ~dt, -dt, -dt)$ time sequence.
In the $(dt, ~dt, -dt, -dt)$ time sequence, only the center layer has $\vec{0}$ parameter and the former (the left) has the positive parameters ($\vec{\theta}$ and $\frac{\vec{\theta}}{2}$) and the latter (the right) has the negative parameters ($-\vec{\theta}$ and $-\frac{\vec{\theta}}{2}$).
Since only the center layer vanishes, this time sequence keeps a similar structure to the target circuit.

Finally, SM offers several advantages over ZNE. Unlike ZNE, it bypasses the need for extrapolation, thereby avoiding the heuristic assumptions that are a key limitation of ZNE. Moreover, SM requires only a single test circuit to estimate the noisy observable. In contrast, ZNE typically involves at least two additional circuits to amplify noise, depending on the gate folding scheme. For instance, with a folding schedule of $(1, 3, 5)$, two additional circuits with 3 times and 5 times noise amplification are needed. As a result, SM incurs a constant sampling overhead, making it more resource-efficient.

\section{Observable Measurement}\label{sec:observable_measurement}

We discuss the effect of the QEM methods described in  \sref{sec:error_mitigations} on the time evolutions of the XXZ spin chain model and the staggered magnetization observable measurement.
In the QEM comparison, we adopt two boundary conditions, the open boundary condition (OBC) and the periodic boundary condition (PBC), with $20$ qubits.
After completing the comparison, we apply the best QEM combination to $84$ (PBC) and $104$ qubit (OBC) cases.
In the large cases, we also compare ZNE and SM methods.

A straightforward approach, referred to as the \textit{direct} method in this work, involves computing the time-evolved expectation value of the staggered magnetization $\hat{O}_{M_{st}}$ for $N$ qubits with respect to the N\'eel state, $\langle\psi(t)|\hat{O}_{M_{st}}|\psi(t)\rangle$ where $|\psi(t)\rangle=e^{-i\,H\,t}|\psi\rangle$. The Hamiltonian $H$ represents the Hamiltonian of the XXZ spin chain in this study. Here, $H$ is a $2^{N}\times 2^{N}$ Hermitian matrix that acts on the Hilbert space of dimension $2^{N}$. 
We implement this approach in QuSpin~\cite{quspin}.
This method is the simplest and yields the highest accuracy. However, it becomes computationally inefficient for time-evolving state vectors when $N\ge 30$ qubits, as the Hilbert space's dimensionality increases exponentially with the number of qubits.
For example, the number of qubits $N=60$ requires 16 exabytes of memory allocation in double precision to represent one state vector. Therefore, we turn to the classical approximation method based on matrix product states to calculate the time evolution of state vectors with $N > 20$ qubits.

Matrix product states (MPS) are a widely used method for studying the time evolution of large quantum many-body systems~\cite{Paeckel:2019yjf, Cirac:2020obd}. An MPS consists of a one-dimensional chain of tensors, each associated with a site or particle in the system. The tensors are connected through bond indices, which can take up to $\chi$ values, referred to as the bond dimension.
Although an MPS can represent any quantum state of a many-body system, capturing the full Hilbert space generally requires a bond dimension $\chi$ that grows exponentially with the system size.
We determine the time evolution of the expectation value of staggered magnetization for the N\'eel state, which we denote here as the \textit{MPS-TDVP} method, using approximation method based on time-dependent variational principle (TDVP)~\cite{TDVP-ref-1, TDVP-ref-2}, utilizing the package ITensor~\cite{itensor-1, itensor-2, itensor-TDVP}. 
Time evolution of MPS using the TDVP-based method offers advantages: it can accommodate Hamiltonians with long-range interactions, not just nearest-neighbor ones, and is also computationally more efficient when periodic boundary conditions (PBC) are applied. In this study, we choose $\delta t=0.1$ as the time-step size, maximum allowed error $\epsilon=10^{-12}$ and maximum bond dimension $\chi_{\mathrm{max}}=1000$ for each sweep in the MPS-TDVP method. 

Besides, in our study we consider the time parameter $t$ in an arbitrary unit where $\hbar=1$ and $J_{1}=1$. One can simply restore $t$ in seconds by mapping $t\rightarrow \hbar t/J_{1}$. For a typical value of exchange interaction, $J_{1}\sim O(\mathrm{eV})$ that is associated with magnetic materials, $t$ falls in $O(10^{-15})$ sec, the time-scale of atomic transitions.

This section focuses on the time evolution of the XXZ spin chain. 
Since our goal is to investigate the accuracy of observable measurements in spin chains on superconducting quantum computers, we concentrate on the time evolution of the expectation value of the staggered magnetization, which captures the antiferromagnetic ordering in the system.
The staggered magnetization observable is represented as follows:
\begin{equation}
    \hat{O}_{M_{st}}=\frac{1}{N}\sum_{i=1}^{N}(-1)^{i}S^{z}_{i} .\label{staggered-mag}
\end{equation}

A specific spin configuration of the quantum spin chain can be selected to compute the expectation value of the staggered magnetization, which serves as a measure of antiferromagnetic ordering in the spin states. While there are many possible choices for such spin configurations, we choose the N\'eel state for its simplicity and its ability to capture key features of antiferromagnetic spin chains.
It is defined as,
\begin{equation}
    \notag
    |\psi (0) \rangle = |\uparrow\downarrow\uparrow\downarrow \cdots \cdots \uparrow\downarrow\uparrow\downarrow\rangle\,
\end{equation}
where each $|\uparrow\rangle$ or $|\downarrow\rangle$ represent the spin projection of spin-$1/2$ particle at i-th site along the $z$-axis in spin space.
Accordingly, we compute the time evolution of the expectation value of the staggered magnetization for the N\'eel state under the given Hamiltonian using IBM's superconducting quantum computers, and validate the results against state-of-the-art classical numerical simulations.

\begin{figure}[h!]
    \centering
    \subfloat[Open Boundary Condition\label{fig:qiskit_N=20:OBC}]{
    \includegraphics[width=0.48\textwidth]{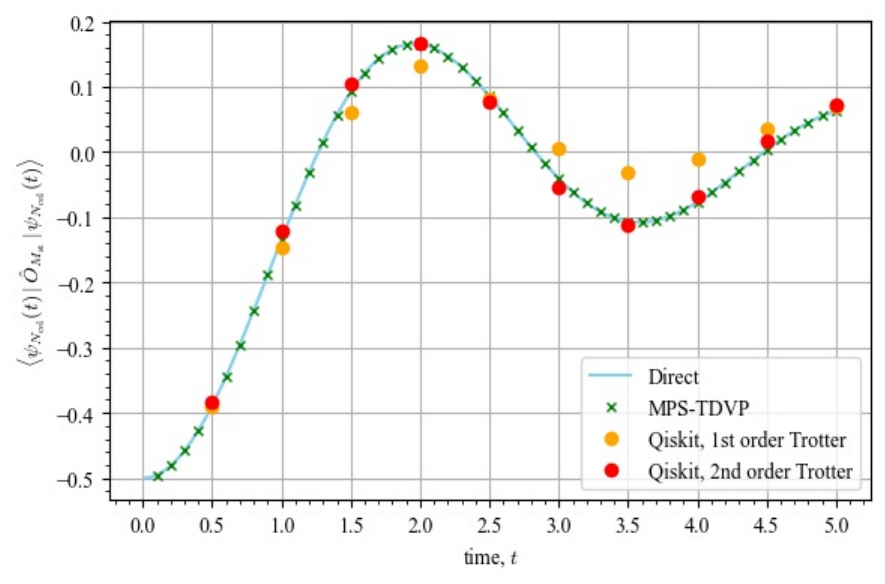}
    }
    \subfloat[Periodic Boundary Condition\label{fig:qiskit_N=20:PBC}]{
    \includegraphics[width=0.48\textwidth]{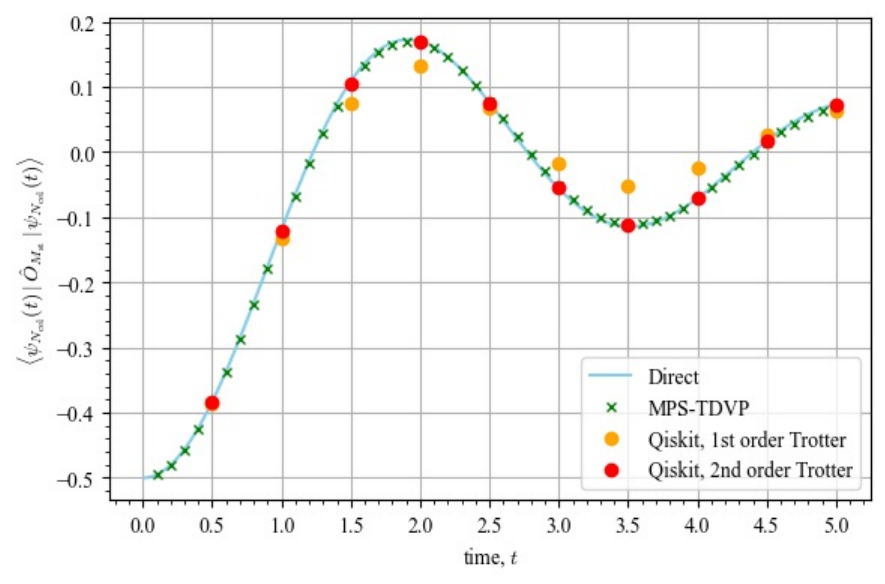}
    }
    \caption{Qiskit simulation of the observable measurement after time evolution of the XXZ model with open and periodic boundary conditions. The Qiskit implementations of the first-order Trotterization and the second-order Trotterization are compared with the direct computation by QuSpin. The MPS-TDVP is plotted to compare its accuracy with Quspin. 
}
    \label{fig:qiskit_N=20}
\end{figure}

As a first step, we compare the first-order Trotterization and the second-order Trotterization in \fref{fig:qiskit_N=20}.
A p-th order Trotterization has $O (t (\delta t)^p)$ error where $t$ and $r$ are the time and the Trotter steps, respectively, and $\delta t = t/r$ \cite{childs2021theory}.
Hence, a fixed time and the Trotter number at the time, a higher order Trotterization has a smaller Trotterization error when $\delta t < 1$.
\Fref{fig:qiskit_N=20} compares the first-order and the second-order Trotterizations with $\delta t = 0.5$ with open boundary condition (OBC) and periodic boundary condition (PBC).
The second-order Trotterization shows a good agreement with the \textit{direct} method, while the first-order Trotterization shows notable disagreement with the \textit{direct} method in both boundary conditions after time $1.0$.
The results show that the second-order Trotterization with $\delta t = 0.5$ is good enough for the observable measurement.
Therefore, we adopt the optimized second-order Trotterization \cite{chowdhury2024enhancing} with $\delta t = 0.5$ and $100,000$ shots in the following experiments on quantum computers.
Also, we compare the \textit{MPS-TDVP} with the direct method to validate its accuracy. As shown in \fref{fig:qiskit_N=20}, \textit{MPS-TDVP} also shows a good agreement with the direct method in both boundary conditions.
Hence, the \textit{MPS-TDVP} is used as the benchmark for $N=84$ (PBC) and $104$ (OBC) cases.

\subsection{Comparison of TREX, DD, and PT}

In this comparison, we measured the time evolution of the expectation value of the staggered magnetization observable with N\'eel's initial state under the XXZ spin chain model of $20$ qubtis using the optimized second-order Trotterization on IBM quantum computer, $\texttt{ibm\_yonsei}$, Eagle r3 quantum processor with $127$ qubits.
Each case on $\texttt{ibm\_yonsei}$ is repeated ten times with 100,000 shots.

In order to systematically benchmark the performance of various QEM techniques on $20$-qubit systems ($N=20$) with OBC and PBC, we conducted a series of experiments under configurations summarized in \tref{tab:qem_configurations}. These include TREX, DD, and PT applied in combinations. The baseline "NOQEM" case serves as a reference for the effect of the QEM technique.

\begin{table}[h!]
\centering
\caption{\label{tab:qem_configurations}Summary of experimental configurations and QEM techniques for system size 20 and boundary conditions.}
\vspace{2mm}
\begin{tabular}{c c p{10cm}}
\br
\textbf{BC} & \textbf{Qubit Count (N)} & \textbf{QEM Techniques Applied} \\
\hline
\hline
OBC & 20 & 1. NOQEM; 2. TREX; 3. TREX + DD; 4. TREX + PT; 5. TREX + DD + PT; \\
\hline
PBC & 20 & 1. NOQEM; 2. TREX; 3. TREX + DD; 4. TREX + PT; 5. TREX + DD + PT; \\
\br 
\end{tabular}

\end{table}

\begin{figure}[h!]
    \centering
    \subfloat[Staggered Magnetization \label{fig:real_N=20:OBC}]{    
    \includegraphics[width=0.48\textwidth]{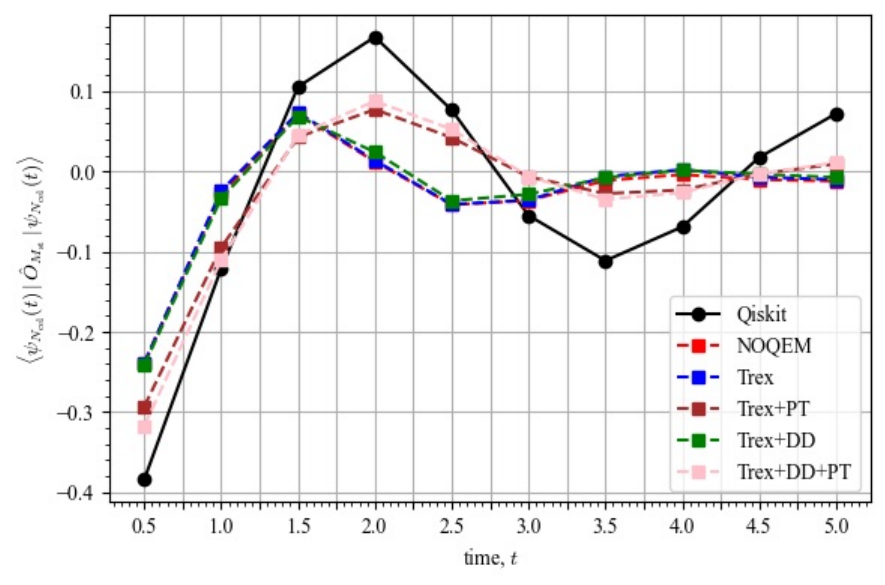}}
    \subfloat[Absolute Error \label{fig:real_N=20_error:OBC}]{    
    \includegraphics[width=0.48\textwidth]{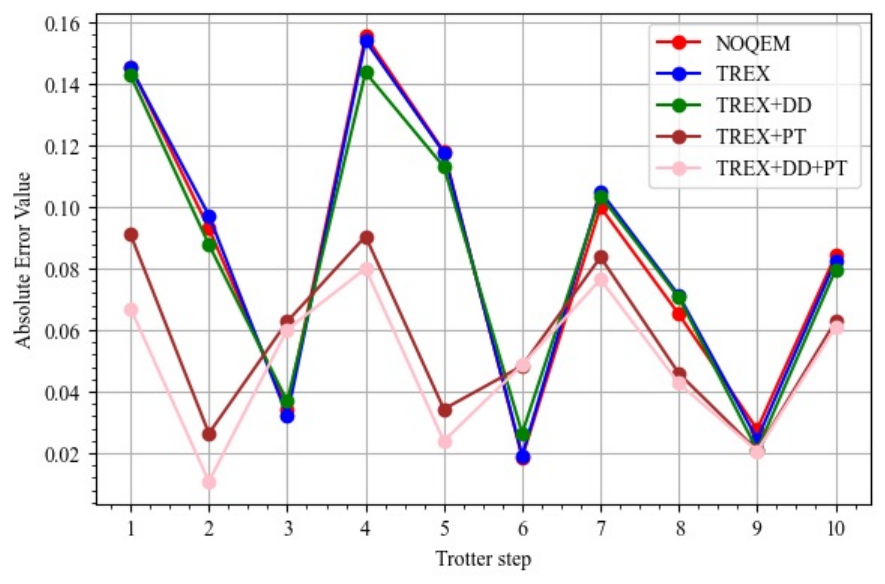}
    }
\caption{
TREX, DD, and PT Comparison with OBC and $N=20$. 
`}
\label{fig:comparison_DDPT_N20_OBC}
\end{figure}

\begin{figure}[h!]
    \centering
    \subfloat[Staggered Magnetization \label{fig:real_N=20:PBC}]{    
    \includegraphics[width=0.48\textwidth]{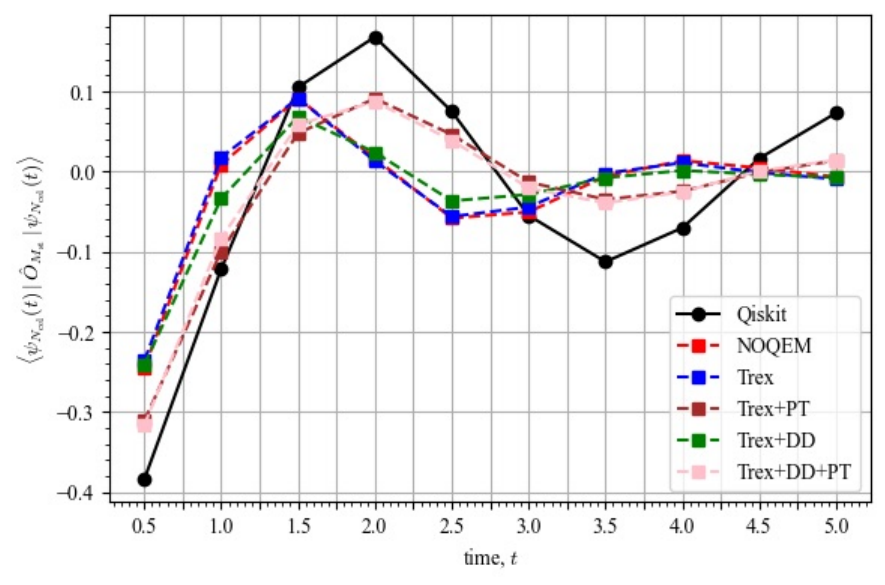}}
    \subfloat[Absolute Error \label{fig:real_N=20_error:PBC}]{    
    \includegraphics[width=0.48\textwidth]{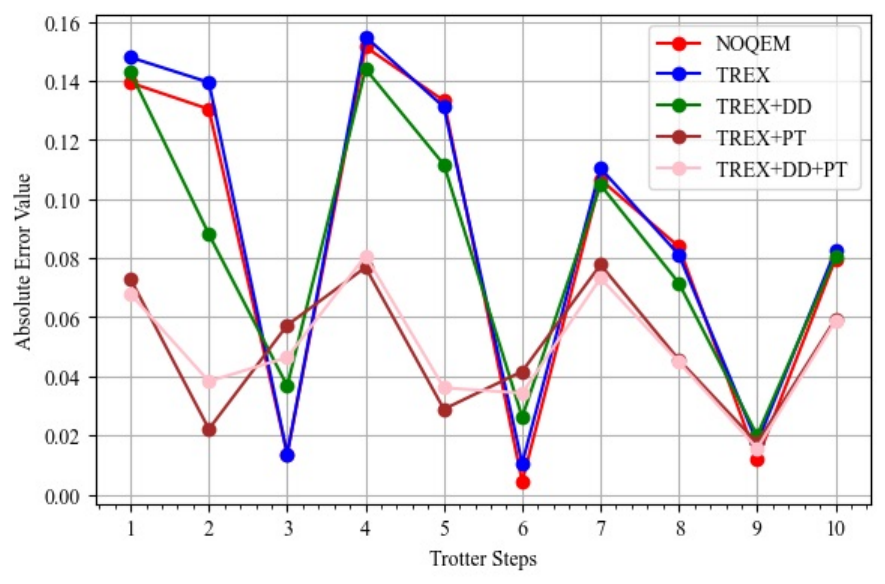}
    }
\caption{
TREX, DD, and PT Comparison with PBC and $N=20$.
}
\label{fig:comparison_DDPT_N20_PBC}
\end{figure}

The comparison results are represented in \fref{fig:comparison_DDPT_N20_OBC},  \fref{fig:comparison_DDPT_N20_PBC}, and Table \ref{tab:qem_observable_comparison} for the OBC and the PBC. 
\Fref{fig:real_N=20:OBC} and \fref{fig:real_N=20:PBC} show the time evolution of the observable $\langle \psi_{\mathrm{ex}}(t) | \mathcal{O}_z | \psi_{\mathrm{ex}}(t) \rangle,$ with the Qiskit reference (black line) compared against results from NOQEM, TREX, TREX+DD, TREX+PT, and TREX+DD+PT on OBC and PBC, respectively.
\Fref{fig:real_N=20_error:OBC} and \fref{fig:real_N=20_error:PBC} present the absolute error rate compared against the Qiskit reference with OBC and PBC, respectively.

\begin{figure}[h!]
    \centering
    \includegraphics[width=0.85\textwidth]{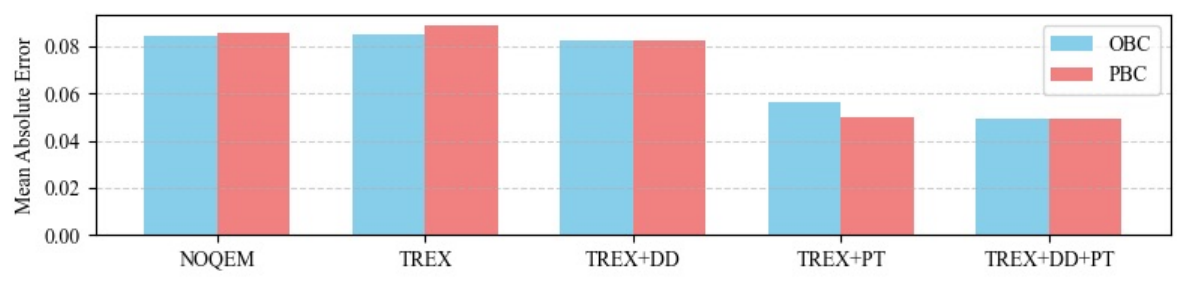}
    \caption{This bar chart presents the error rates of various Quantum Error Mitigation (QEM) strategies under Open Boundary Conditions (OBC, blue) and Periodic Boundary Conditions (PBC, red) with $N=20$. }
    \label{fig:N=20_Errorcombined}
\end{figure}

Among the methods, both TREX+DD+PT and TREX+PT closely reproduce the overall shape of the reference curve, aside from amplitude discrepancies in both OBC and PBC cases, with TREX+DD+PT showing slightly higher accuracy under both boundary conditions.
In contrast, results from NOQEM, TREX, and TREX+DD show notable discrepancies with the reference curve at the early time and decay of the observables to zero after time $1.5$ (Trotter step $3$), likely due to noise-induced distortions by the device error. The quick decay of the observables of NOQEM, TREX, and TREX+DD causes them to align with the reference curve at specific time points incidentally, where the reference values have near-zero values, resulting in temporarily lower absolute errors despite overall poorer performance.
This is observed at Trotter steps 3 and 6 in \fref{fig:real_N=20_error:OBC} and \fref{fig:real_N=20_error:PBC}, which presents the absolute error at each Trotter step. To mitigate such anomalies and assess overall performance more robustly, the average error values are reported separately in \fref{fig:N=20_Errorcombined}, clearly indicating the superior accuracy and stability of the TREX+DD+PT strategy across the entire time evolution under both boundary conditions.

The comparison between TREX+DD and TREX+PT in \fref{fig:N=20_Errorcombined} shows that the contribution by DD is slight, even though the performance of DD is practically validated in several experiments \cite{coote2025resource, tong2024empirical, ezzell2023dynamical, liu2013noise}.
We conjecture that DD has limited impact in our implementation because the circuit (see \fref{fig:circuit_XXX_2nd_opt}) follows a brick-wall structure, with each brick constructed as described in equation~\eref{eq:circuit_XYZ}.
In other words, because our circuit implementation is already dense, there is little room to insert additional DD scheduling.
This conjecture is also supported by the PBC case (red histograms) in \fref{fig:N=20_Errorcombined}. Since the PBC implementation has the brick implementation between the boundary qubits (qubit $0$ and $19$) in the odd layers, the PBC has less room for the DD scheduling.
Hence, the difference between TREX+PT and TREX+DD+PT is negligible.
In other words, DD scheduling is unnecessary for the PBC case due to the lack of available space for its insertion.

\subsection{Comparison of ZNE and SM}

\begin{figure}[h!]
    \centering
    \subfloat[Staggered Magnetization]{    
    \includegraphics[width=0.48\textwidth]{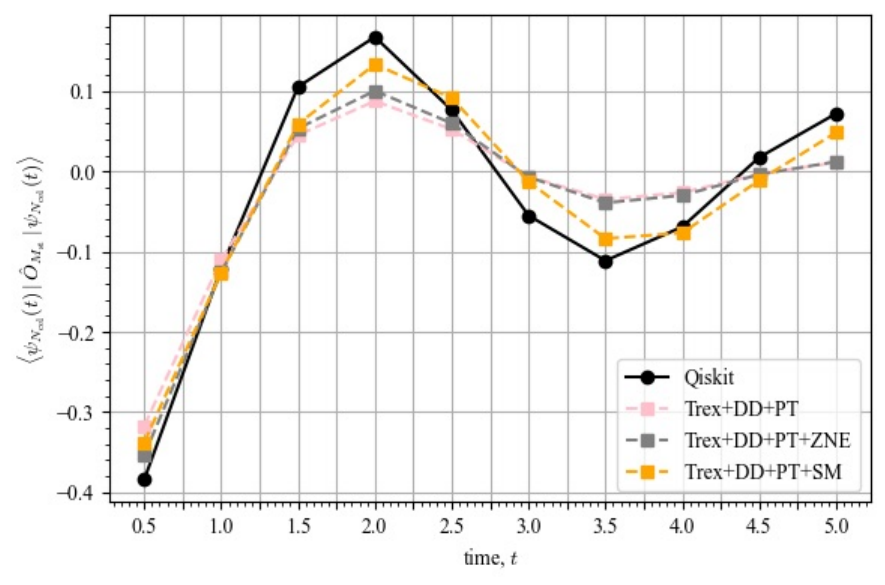}}
    \subfloat[Absolute Error]{
    \includegraphics[width=0.48\textwidth]{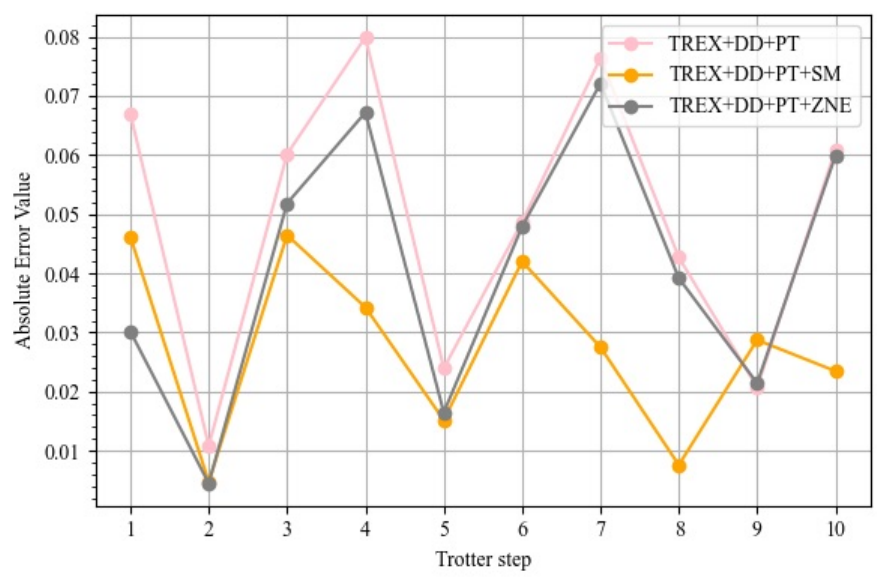}}
    \caption{Comparison of ZNE and SM with OBC and $N=20$. 
    }
    \label{fig:comparison_ZNESM_N20:OBC}
\end{figure}

\begin{figure}[h!]
    \centering
    \subfloat[Staggered Magnetization]{    
    \includegraphics[width=0.48\textwidth]{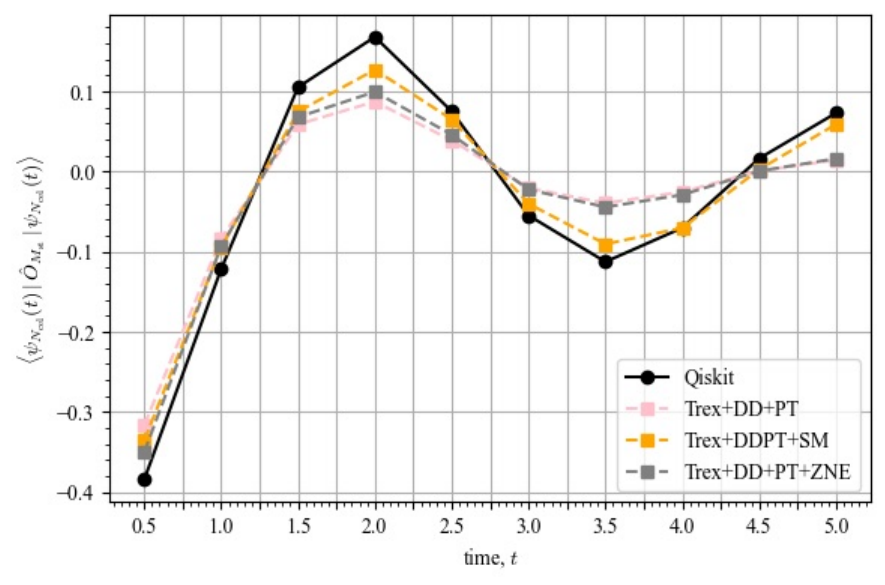}}
    \subfloat[Absolute Error]{
    \includegraphics[width=0.48\textwidth]{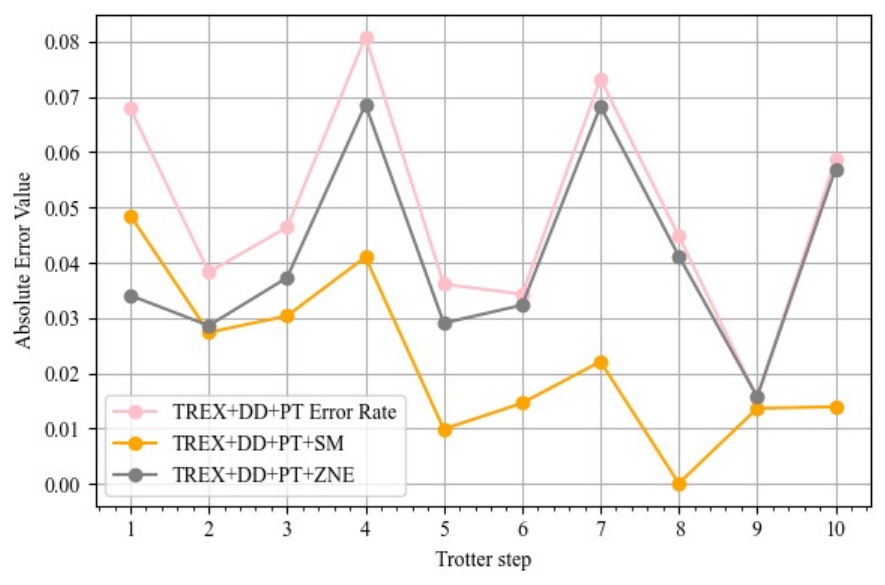}}
    \caption{Comparison of ZNE and SM with PBC and $N=20$
    }
    \label{fig:comparison_ZNESM_N20:PBC}
\end{figure}

This section extends the QEM comparison to the ZNE and the SM on $\texttt{ibm\_yonsei}$.
Building on the prior analysis showing that combining TREX, DD, and PT provides the highest accuracy, we now incorporate and compare ZNE and SM within the TREX+DD+PT framework.
The Qiskit results shown in \fref{fig:comparison_ZNESM_N20:OBC} and \fref{fig:comparison_ZNESM_N20:PBC} serve as the benchmark for evaluating the absolute error.

\begin{figure}[h!]
    \centering
    \includegraphics[width=0.75\textwidth]{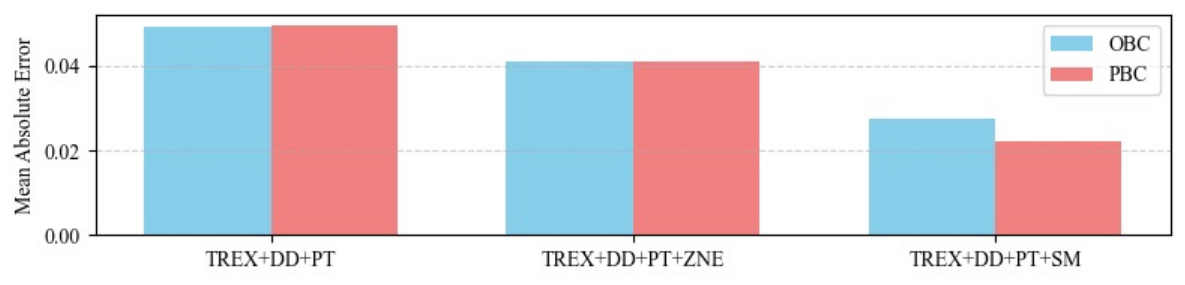}
    \caption{Comparison of the mean absolute errors for the ZNE and the SM with OBC and PBC when $N=20$. The TREX+DD+PT result is presented again to compare with the ZNE and the SM.}
    \label{fig:N=20_ZNE_and_SM}
\end{figure}

As shown in \fref{fig:comparison_ZNESM_N20:OBC} (OBC) and \fref{fig:comparison_ZNESM_N20:PBC} (PBC), the SM method is more accurate at all the time steps except for the time step $9$ of the OBC case (\fref{fig:comparison_ZNESM_N20:OBC}).
At that point, the true value has a near-zero value, and the ZNE results decay to zero after time $3.0$ (Trotter step $6$), likely due to noise-induced distortions by the device error.
Due to the two effects combined incidentally, the ZNE result shows better absolute error than the SM result at the Trotter step $9$ of the OBC case.
Except for that point, the SM method shows more accurate results than the ZNE method, even though the ZNE always shows a better accuracy than the TREX+DD+PT method, which is the baseline of this comparison.

The corresponding average absolute error rates of these results are summarized in \fref{fig:N=20_ZNE_and_SM} and \tref{tab:qem_mae_conversion_extended}. Specifically, we compare the performance of ZNE and SM, both applied on top of the TREX+DD+PT baseline.
The average absolute error value with ZNE is $0.04106$ and $0.04123$ on OBC and PBC, respectively, whereas the SM has $0.02760$ and $0.02218$ on OBC and PBC, respectively.
The SM achieves an average absolute error reduction of $32.8\%$ for OBC and $46.2\%$ for PBC, respectively.
Interestingly, applying SM yields better results than the full combination with ZNE, suggesting that SM’s circuit-aware mitigation is more effective than ZNE’s heuristic extrapolation under both boundary conditions.

\begin{figure}[h!]
    \centering
    \subfloat[Staggered Magnetization \label{fig:real_N104_SMZNE:OBC:Obs}]{\includegraphics[width=0.48\textwidth]{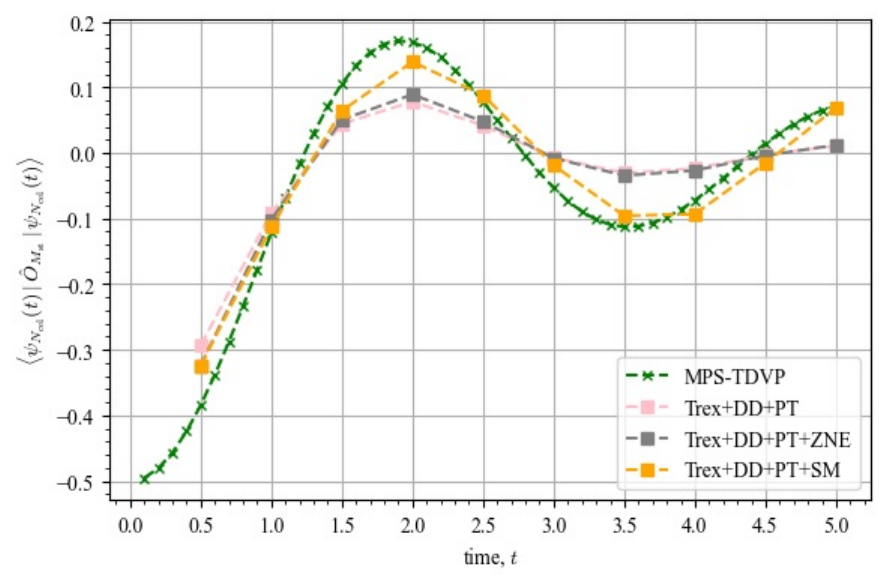}}
    \subfloat[Absolute Error \label{fig:real_N104_SMZNE:OBC:AE}]
    {\includegraphics[width=0.48\textwidth]{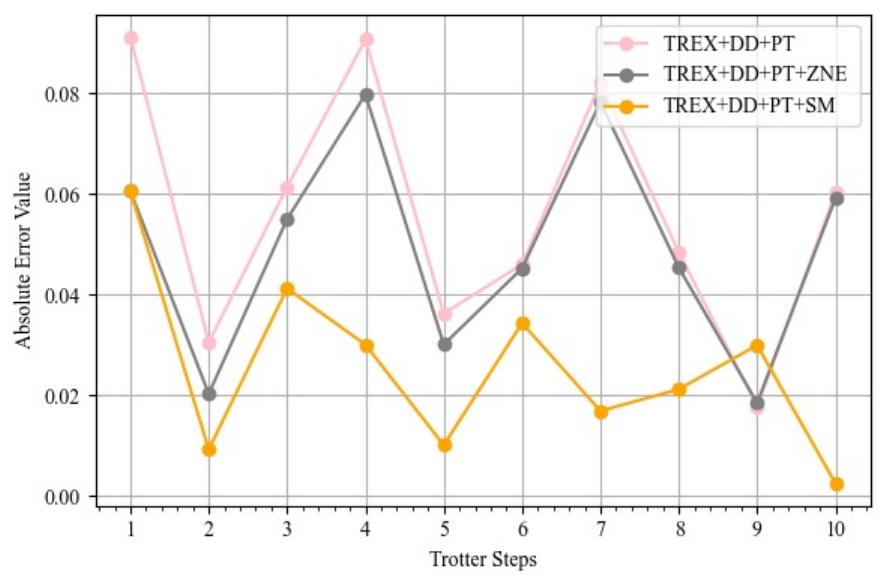}}
    \caption{Comparison of ZNE and SM with OBC and $N=104$.}
    \label{fig:N104_OBC}
\end{figure}

\begin{figure}[h!]
    \centering
    \subfloat[Staggered Magnetization \label{fig:real_N=84_SMZNE:PBC:Obs}]{\includegraphics[width=0.48\textwidth]{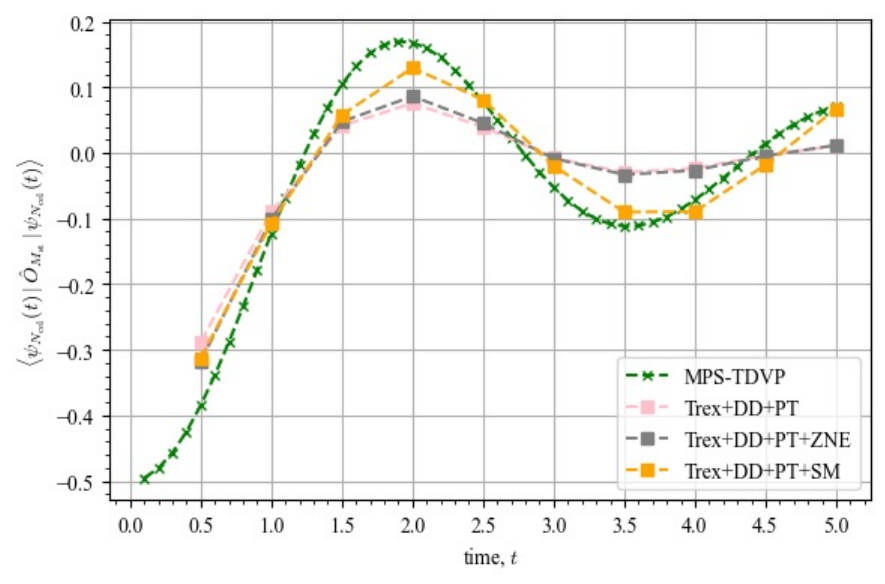}}
    \subfloat[Absolute Error \label{fig:real_N=84_SMZNE:PBC:AE}]
    {\includegraphics[width=0.48\textwidth]{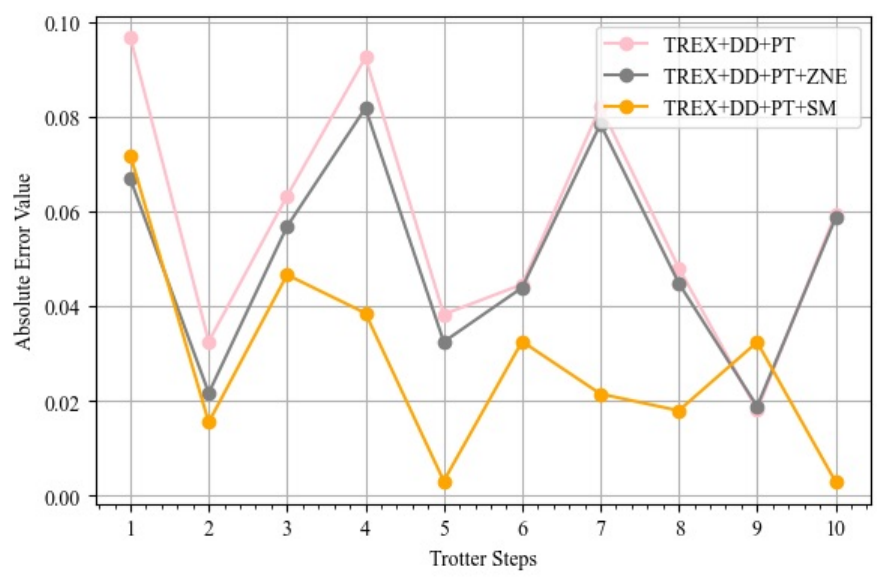}}
    \caption{Comparison of ZNE and SM with PBC and $N=84$.}
    \label{fig:N84_PBC}
\end{figure}

We extend the system size to $N=104$ for OBC and $N=84$ for PBC.
The PBC system is limited to 84 qubits due to the requirement of connecting the first and last qubits. For the ZNE, we apply the same extrapolation fitting curve used in the $N=20$ cases.
As a benchmark for the true value, we use the \textit{MPS-TDVP} method, as outlined earlier in this section since the direct simulation is not feasible for systems of these sizes. The accuracy of the \textit{MPS-TDVP} is validated in \fref{fig:qiskit_N=20}.
These simulations use more than $3,200$ and $2,600$ two-qubit gates on the $OBC$ and the $PBC$, respectively, as listed in \tref{tab:circuit_stat}.

\begin{figure}[t!]
    \centering
    \includegraphics[width=0.75\textwidth]{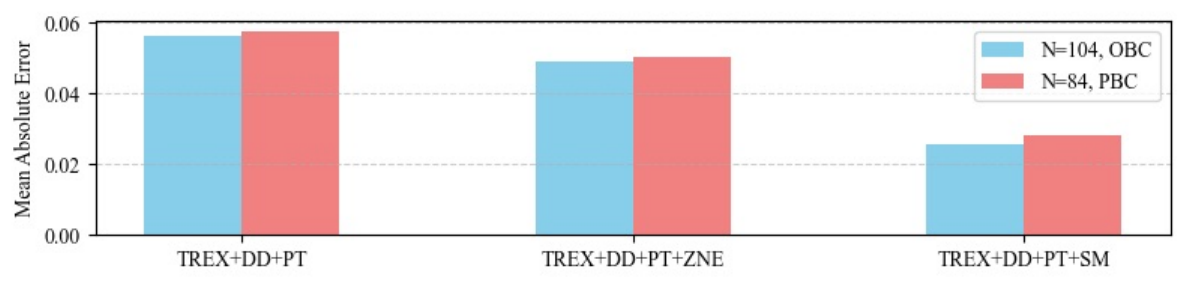}\hspace{1mm}
\caption{Comparison of the mean absolute errors of the TREX+DD+PT, the ZNE, and the SM with OBC ($N=104$) and PBC ($N=84$).}
    \label{fig:N=104andN=84_errorcombined}
\end{figure}

\Fref{fig:N104_OBC} and \fref{fig:N84_PBC} show that the SM method aligns much more closely with the benchmark, represented by green x marks in the figures, than ZNE, even at the larger system sizes of $N=104$ and $N=84$.
Note that the more accurate results of the ZNE at the Trotter step $9$ in both cases are caused incidentally, as explained in the cases of 20-qubit systems (\fref{fig:comparison_ZNESM_N20:OBC} and \fref{fig:comparison_ZNESM_N20:PBC}).
The average absolute errors for each approach are presented in  \fref{fig:N=104andN=84_errorcombined} and ~\tref{tab:qem_mae_conversion_extended}.
The average absolute error value with ZNE is $0.04917$ and $0.05044$ on OBC and PBC, respectively, whereas the SM has $0.02556$ and $0.02832$ on OBC and PBC, respectively.
The SM achieves an average absolute error reduction of $48.0\%$ on OBC and $43.9\%$ on PBC compared to ZNE.
Furthermore, the average absolute errors of ZNE increase by $19.8\%$ and $22.3\%$ compared to the 20-qubit cases for OBC and PBC, respectively.

In conclusion, the SM yields the lowest mean absolute error across all tested system sizes and boundary conditions, clearly outperforming both ZNE and the baseline (TREX+DD+PT) among the three methods.
When comparing the 20-qubit systems with the larger systems (104 and 84 qubits), the SM method maintains similar absolute errors, whereas the ZNE method exhibits increased absolute errors in the larger systems. This likely stems from the fact that the ZNE extrapolation curve, derived from the 20-qubit systems, does not accurately capture the error characteristics of the larger systems. In contrast, the SM method more effectively estimates the noise factors in both small and large-scale systems since SM leverages the circuit's inherent noise characteristics and preserves robustness without increasing circuit depth. These features make SM a promising approach for scalable and reliable quantum simulations of Trotterization on near-term noisy quantum devices.
However, the SM technique is particularly effective when the true observable can be readily estimated, such as in Trotterization, while ZNE can be applied to more general cases.

\section{Entanglement Entropy Measurement}  
\label{sec:entropy}

In this section, we explore measuring entanglement entropy after the time evolution of the Hamiltonian on IBM quantum computers.
Measuring entanglement entropy is important because it provides deep insight into the quantum structure and correlations of a system. Entanglement entropy quantifies how strongly different parts of a quantum system are entangled. Unlike classical correlations, entanglement captures nonlocal quantum correlations that are fundamental to quantum mechanics.
In particular, in many-body physics, entanglement entropy helps distinguish different quantum phases, such as topological phases, where traditional order parameters fail, and critical phases, where the entropy scales logarithmically with subsystem size.

Despite measuring entanglement entropy being a powerful tool to study quantum systems, it is quite challenging on real quantum computers.
Even though we can reconstruct the density matrix of the given system by quantum state tomography (QST), QST is not practical except for very small system sizes because it needs exponential operations with respect to the system size.
Hence, we briefly describe a practical entanglement entropy measurement method. When we have a system of $N$ qubits in a pure state $|\psi\rangle$, with its density matrix, $\rho = |\psi\rangle\langle\psi|$ and we partition this system into two subsystems consisting of  $L$ and $M = N - L$ qubits, the entanglement entropy, specifically the  R\'enyi entropy of order $n$ associated with the $L$-qubit subsystem, is defined as
\begin{equation}
S^{(n)}_{L} = \frac{1}{1-n} \log \left[ \mathrm{Tr}(\rho_{L}^{n}) \right],
\label{eq:n-Renyi entropy}
\end{equation}
where $\rho_{L}$ is the reduced density matrix for the subsystem with  $L$ qubits. This reduced density matrix is obtained by tracing out the subsystem of $M = N - L$ qubits, such that $\rho_{L} = \mathrm{Tr}_{M} \rho$. 
In this study, we focus on the R\'enyi entropy of order $2$ (simply referred to as R\'enyi entropy from now on).
Note that the R\'enyi entropy of order $n$ converges to the von Neumann entropy :
\begin{equation}
\notag S^{vN}_{L} = -\mathrm{Tr} \left( \rho_{L} \log \rho_{L} \right)
\label{eq:von neumann entropy}
\end{equation}
as $n$ approaches 1.

One well-known practical quantum algorithm for the entanglement measurement is the swap-based many-body interference (SWAP-MBI) protocol~\cite{Buhrman:2001rma, Horodecki-swap, Ekert-swap} based on the swap test, which is presented in the quantum circuit.
When we have an $N$ qubit system, the protocol needs $2N+1$ qubits because it has two copies of the system and one ancilla (control) qubit.
Even though the SWAP-MBI protocol is much more efficient than the QST, it is still limited when the connectivity between qubits is limited, such as in superconducting quantum computers, because the ancilla qubit needs to be connected to all other qubits. 
Although an efficient implementation of the SWAP-MBI on superconducting quantum computers is proposed \cite{chowdhury2024capturing, chowdhury2025first} on a small system, it is not practical on larger systems.
Hence, we adopt the randomized measurement (RM) protocol.

\subsection{Randomized Measurement Protocol}\label{sec:randomized-protocol}

\begin{figure}[t!]
\[
\Qcircuit @C=1.0em @R=0.3em @!R{
\lstick{q_{0}: }   & \qw & \multigate{7}{ \makecell{\mathrm{Trotter}\\\mathrm{Steps}} } & \qw & \gate{SU(2)} & \meter &\qw\\
\lstick{q_{1}: }   & \qw & \ghost{ \makecell{\mathrm{Trotter}\\\mathrm{Steps}} }        & \qw & \gate{SU(2)} & \meter & \qw\\
\vdots & & & \vdots & &  \vdots\\
\lstick{q_{L-1}: } & \qw & \ghost{ \makecell{\mathrm{Trotter}\\\mathrm{Steps}} }        & \qw & \gate{SU(2)} & \meter & \qw \\
\lstick{q_{L}: } & \qw & \ghost{ \makecell{\mathrm{Trotter}\\\mathrm{Steps}} }        & \qw & \qw & \qw & \qw \\
\vdots & & & \vdots & &  \vdots\\
\lstick{q_{N-2}: } & \qw & \ghost{ \makecell{\mathrm{Trotter}\\\mathrm{Steps}} }        & \qw & \qw & \qw & \qw \\
\lstick{q_{N-1}: } & \qw & \ghost{ \makecell{\mathrm{Trotter}\\\mathrm{Steps}} }        & \qw & \qw & \qw    & \qw
}
\]
\caption{The randomized measurement (RM) protocol, where the Trotter steps prepare the quantum state. Then randomized measurements on a subsystem A (system size $L$; the upper-subsystem as an example) are performed by applying $N_{U}$ product of local random unitaries $\hat{U}_{a}=\otimes_{i=1}^{L}U^{(2)}_{i}$ where each $U^{(2)}_{i}$ is sampled from CUE. 
}
\label{fig:randomized-protocol}
\end{figure}
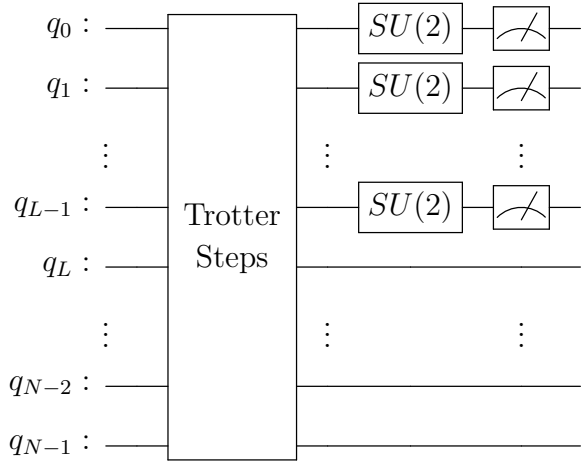

The RM protocol~\cite{Brydges-randomized, Enk-Beenakker, Elben-1, Vermersch-1, Elben-2, Rath, Elben-3} leverages the insight that the R\'enyi entropy of a quantum system can be inferred from statistical correlations among measurement outcomes in randomly chosen bases. In our case, we consider a pure quantum state 
$\rho$ of $N$ qubits generated by a random unitary circuit. Consequently, the purity and R\'enyi entropy are evaluated for the reduced density matrix corresponding to a subsystem $A$ consisting of $L<N$ qubits. A schematic quantum circuit illustrating the RM protocol is shown in \fref{fig:randomized-protocol}. The experimental realization of this protocol involves several steps, which we outline below.

First, the quantum state $\rho$ is generated by applying the Trotterization steps of $N$ qubits, resulting in a highly entangled state. Our goal is to measure the purity and R\'enyi entropy of a subsystem $A$ consisting of 
$L$ qubits.
We apply a product of single-qubit unitary operators where each of them acts on the $i$-th qubit of the $A$ subsystem, i.e., $i\in A$, as 
\begin{equation}
\notag
 \hat{U}_{a}=U^{(2)}_{1}\otimes U^{(2)}_{2}\otimes...\otimes U^{(2)}_{L} .
 \label{eq:single-qubit-local-unitary}
\end{equation}

After that, each of the single-qubit unitary $U^{(2)}_{i}$ is independently sampled from the circular unitary ensemble (CUE) of the $SU(2)$. Afterward, the qubits are measured on a computational basis (i.e., the $Z$-basis).
For each $\hat{U}_{a}$, repeated measurements are performed to gather statistics (commonly referred to as shots), allowing for the estimation of occupation probabilities, $P_{\hat{U}_{a}}=\mathrm{Tr}[\hat{U}_{a}\rho\hat{U}^{\dagger}_{a}|j_{L}\rangle\langle j_{L}|]$ of computational basis states $|j_{L}\rangle = |s_{1},s_{2},...,s_{L}\rangle$ with $s_{i}=0,1$. Note that $\hat{U}_{a}$ acts only on the subspace of $L$ qubits. Afterward, the entire process is repeated for $N_{U}$ different randomly drawn instances of $\hat{U}_{a}$.

Once the set of outcome probabilities $P_{\hat{U}_{a}}(j_{L})$ of the computational basis states $|j_{L}\rangle$ for one instance of random unitaries, $\hat{U}_{a}$, the following quantity is computed:
\begin{equation}
\notag
    X_{a} = 2^{L}\sum_{j_{L},j'_{L}}(-2)^{-D[j_{L},j'_{L}]}P_{\hat{U}_{a}}(j_{L})P_{\hat{U}_{a}}(j'_{L})
\end{equation}
where, $j_{L}$ and $j'_{L}$ are different bitstring outcomes for the $L$ qubits, and $P_{\hat{U}_{a}}(j_{L}),\,P_{\hat{U}_{a}}(j'_{L})$ are their corresponding outcome probabilities for the corresponding $\hat{U}_{a}$. 
Additionally, $D[j_{L},j'_{L}]$ is the Hamming distance between bitstrings $j_{L}=s_{1}s_{2}...s_{L}$ and $j'_{L}=s'_{1}s'_{2}...s'_{L}$, measuring how different two bitstrings are, i. e., $D[j_{L},j'_{L}]\equiv \#\{i\in A|s_i\neq s'_i\}$. 
Consequently the ensemble average of $X_{a}$, denoted by $\overline{X}$,
\begin{equation}
    \overline{X} = \frac{1}{N_{U}}\sum_{a=1}^{N_{U}}X_{a}
    \label{eq:RM-estimated-purity}
\end{equation}
is nothing but the second-order cross-correlations across the ensemble of discrete $N_{U}$ random unitaries $\hat{U}_{a}$, and provides the estimation of the purity $\mathrm{Tr}(\rho_{A}^{2})$ associated with a subsystem of $L$ qubits, $\{q_{i_{1}},...,q_{i_{L}}\}$ where indices $i_{1},..., i_{L}$ are chosen from $[0, N-1]$ in a system of $N$ qubits. Finally, the R\'enyi entropy (of order 2) is 
\begin{equation}
\notag
    S^{(2)} = -\mathrm{log}\overline{X} .
\end{equation}

Note that the RM protocol can be recast as the swap-based many-body interference protocol where the swap operator effectively acts on the two virtual copies of $\rho_{A}$ \cite{Elben-2}.

\subsection{Quantum Multi-Programming}

A central challenge of the RM protocol lies in its dependence on an ensemble of locally applied Haar-random unitaries to suppress variance in entropy estimation. In our experiments, we implement 60 distinct instances of the RM protocol circuits. Additionally, to mitigate errors, we apply TREX, DD, PT, and ZNE. Since applying ZNE and PT repeat each circuit three times (1, 3, 5 gate folding) and ten times for PT (10 PT pair sampling), respectively, in the experimental setting, the total number of circuits executed per the entanglment measurement is 1,800 ($ 60 \times 3 \times 10$).

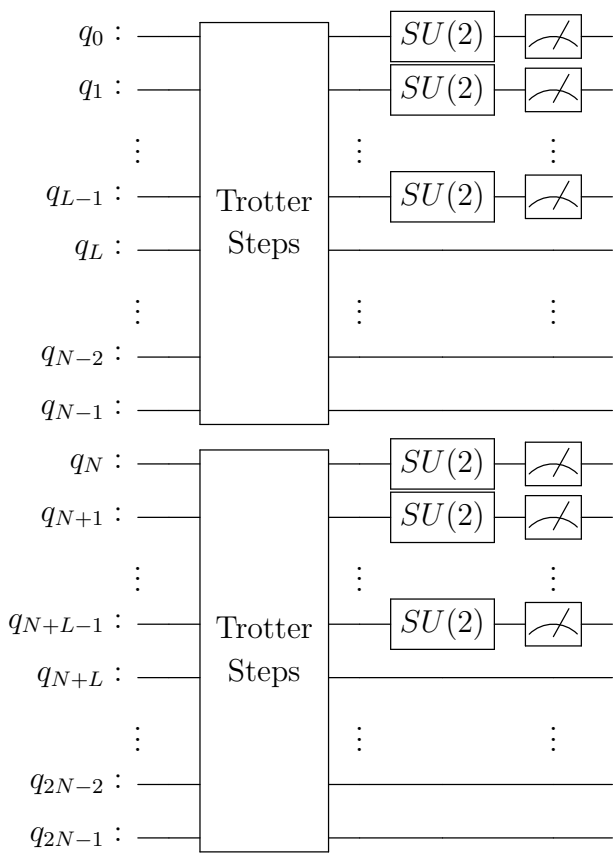
\begin{figure}[t!]
\[
\Qcircuit @C=1.0em @R=0.1em @!R{
\lstick{q_{0}: }   & \qw & \multigate{7}{ \makecell{\mathrm{Trotter}\\\mathrm{Steps}} } & \qw & \gate{SU(2)} & \meter &\qw\\
\lstick{q_{1}: }   & \qw & \ghost{ \makecell{\mathrm{Trotter}\\\mathrm{Steps}} }        & \qw & \gate{SU(2)} & \meter & \qw\\
\vdots & & & \vdots & &  \vdots\\
\lstick{q_{L-1}: } & \qw & \ghost{ \makecell{\mathrm{Trotter}\\\mathrm{Steps}} }        & \qw & \gate{SU(2)} & \meter & \qw \\
\lstick{q_{L}: } & \qw & \ghost{ \makecell{\mathrm{Trotter}\\\mathrm{Steps}} }        & \qw & \qw & \qw & \qw \\
\vdots & & & \vdots & &  \vdots\\
\lstick{q_{N-2}: } & \qw & \ghost{ \makecell{\mathrm{Trotter}\\\mathrm{Steps}} }        & \qw & \qw & \qw & \qw \\
\lstick{q_{N-1}: } & \qw & \ghost{ \makecell{\mathrm{Trotter}\\\mathrm{Steps}} }        & \qw & \qw & \qw    & \qw \\
\lstick{q_{N}: }   & \qw & \multigate{7}{ \makecell{\mathrm{Trotter}\\\mathrm{Steps}} } & \qw & \gate{SU(2)} & \meter &\qw\\
\lstick{q_{N+1}: }   & \qw & \ghost{ \makecell{\mathrm{Trotter}\\\mathrm{Steps}} }        & \qw & \gate{SU(2)} & \meter & \qw\\
\vdots & & & \vdots & &  \vdots\\
\lstick{q_{N+L-1}: } & \qw & \ghost{ \makecell{\mathrm{Trotter}\\\mathrm{Steps}} }        & \qw & \gate{SU(2)} & \meter & \qw \\
\lstick{q_{N+L}: } & \qw & \ghost{ \makecell{\mathrm{Trotter}\\\mathrm{Steps}} }        & \qw & \qw & \qw & \qw \\
\vdots & & & \vdots & &  \vdots\\
\lstick{q_{2N-2}: } & \qw & \ghost{ \makecell{\mathrm{Trotter}\\\mathrm{Steps}} }        & \qw & \qw & \qw & \qw \\
\lstick{q_{2N-1}: } & \qw & \ghost{ \makecell{\mathrm{Trotter}\\\mathrm{Steps}} }        & \qw & \qw & \qw    & \qw
}
\]
\caption{The Quantum Multi-Programming (QMP) implementation of the randomized measurement (RM) protocol in \fref{fig:randomized-protocol}. 
}
\label{fig:randomized-protocol_QMP}
\end{figure}

To make efficient use of the IBMQ system, we employ Quantum Multi-Programming (QMP) to parallelize circuit execution. QMP enables the simultaneous execution of multiple quantum circuits, irrespective of their types or depths.
The effectiveness of QMP has been demonstrated across a range of quantum algorithms, including Grover's search, quantum amplitude estimation, the quantum support vector machine, and measuring R\'enyi entropy \cite{park2023quantum, rao2024quantum, baker2024parallel, chowdhury2024capturing, chowdhury2025first}.
Despite its efficiency, QMP introduces certain challenges, such as crosstalk between concurrently executed circuits. To address these issues, we adopt the mitigation strategy described in Ref.~\cite{park2023quantum}. In our QMP implementation, we parallelize two circuits per QMP package, inserting one physical idle qubit between them to reduce interference. As a result, to execute the full set of 1,800 circuits, we run 900 parallelized circuit pairs.
The circuit diagram of the QMP implementation is visualized in \fref{fig:randomized-protocol_QMP}.
After executing the QMP circuit, both circuits are measured simultaneously. The measurement outcomes for each circuit are then post-processed following the procedure outlined in Appendix A of Ref. ~\cite{park2023quantum}.

\subsection{Results}
\label{sec:EEM_results}

\begin{figure}[h!]
    \centering
    \includegraphics[width=0.80\textwidth]{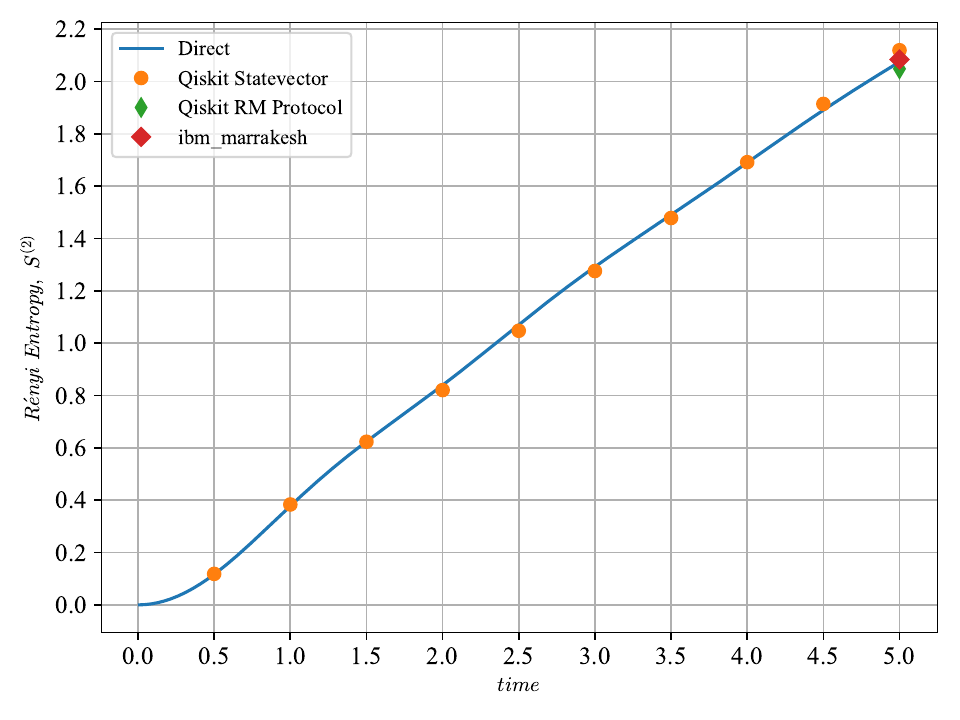} 
\caption{R\'enyi entropy of order 2 after time evolution of the Hamiltonian.}
    \label{fig:EE_S2_N20}
\end{figure}

The measurement results of the entanglement entropy are shown in \fref{fig:EE_S2_N20}.
The Direct results are computed using QuSpin (see \sref{sec:observable_measurement} for details on the method).
The Qiskit Statevector results are obtained using an optimized second-order Trotterization of the Hamiltonian. In this computation, the Rényi entropy is calculated from the statevector after time evolution. The slight discrepancy observed from the fourth time step onward is attributed to Trotterization error.
The Qiskit RM Protocol results are generated using the randomized measurement (RM) protocol described in \sref{sec:randomized-protocol}, applied after time evolution using the same optimized second-order Trotterization. Hence, this result includes both Trotterization and RM protocol errors.
Finally, the \texttt{ibm\_marrakesh} results are obtained from an IBM Quantum device, \texttt{ibm\_marrakesh}, which features a Heron r2 processor with 156 qubits.
The Qiskit RM Protocol and \texttt{ibm\_marrakesh} cases use identical circuits to eliminate variability due to circuit differences. As a result, the \texttt{ibm\_marrakesh} outcome reflects the combined effects of Trotterization error, RM protocol error, and hardware-induced quantum device error. Additionally, the QEM techniques described in \sref{sec:error_mitigations} are applied to the \texttt{ibm\_marrakesh} data.

The Direct, Qiskit Statevector, Qiskit RM Protocol, and \texttt{ibm\_marrakesh} has $2.07465$, $2.11936$, $2.04836$, and $2.08378$, respectively, at time $5.0$.
The Qiskit Statevector has $2.16\%$ relative error with respect to the Direct.
The Qiskit RM Protocol has $3.35\%$ relative error with respect to the Qiskit Statevector.
The \texttt{ibm\_marrakesh} has $1.73\%$ relative error with respect to the Qiskit RM Protocol.
Therefore, these results demonstrate that the entanglement entropy measured on the quantum computer aligns well with the noiseless classical simulation.

\section{Conclusion and outlook}\label{sec:conclusion}

In this study, we compare a range of QEM methods, examining their characteristics as well as the advantages and limitations of each.
Building on this comparative analysis, we assess the capability of current noisy quantum computers to measure observables and entanglement entropy—key quantities for gaining deep insights into quantum systems.
We then extend the observable measurements to the quantum utility scale, involving systems with over 100 qubits.

This research primarily addresses the pre- and early-FTQC stages, but the QEM methods discussed in this paper will continue to offer utility as FTQC systems mature because the mature FTQC is expected to exhibit LER in the range of $10^{-12}$ to $10^{-15}$, which are substantially higher than the machine epsilon error (approximately $2.22 \times 10^{-16}$ in IEEE 754 double precision) found in classical computers.

The randomized measurement protocol is fundamentally constrained by the high number of measurements needed to reduce statistical errors. Even though an advanced method, such as importance sampling \cite{Rath}, has been developed, the number of randomized measurements with a given accuracy is still of the order of $2^{aN}$ with $a \approx 1$ when the system has $N$ qubits \cite{Elben-1, Elben-3, huang2020predicting}.
We will continue our ongoing efforts to develop more robust and scalable quantum algorithms for measuring entanglement entropy.
Despite the limitation of the entanglement entropy, this study will be a significant step forward in the realm of quantum simulation at the utility scale before the fault-tolerant quantum computing era.

This study expands the applicability of noisy quantum computers to more complex scenarios, specifically addressing large-scale time dynamics of Hamiltonians on noisy superconducting quantum hardware.
Looking ahead, our efforts will lay the groundwork for future quantum computing applications, highlighting the quantum advantage over classical methods in simulating complex quantum systems. As we continue to advance the frontiers of quantum computing, our findings add to the growing evidence of its transformative potential in deepening our understanding of quantum phenomena.

\section*{Acknowledgments}
K.Y. is supported by the Brookhaven National Laboratory LDRD \#24-061, \#25-033, and the U.S. Department of Energy, Office of Science, Grants No. DE-SC0012704.
This research was supported by the Institute for Convergence Research and Education in Advanced Technology (iCREATE) and the Institute of Quantum Information Technology (IQIT) at Yonsei University through the use of the Yonsei quantum computing system, and by the APEC-APRU partnership program (B0080429002759) funded by the Ministry of Education, Korea.
This research used quantum computing resources of the Oak Ridge Leadership Computing Facility, which is a DOE Office of Science User Facility supported under Contract DE-AC05-00OR22725. 
This research used resources of the National Energy Research Scientific Computing Center, a DOE Office of Science User Facility supported by the Office of Science of the U.S. Department of Energy under Contract No. DE-AC02-05CH11231 using NERSC award NERSC DDR-ERCAP0028999 and DDR-ERCAP0033558.

\appendix

\section{Experimental Data}

\begin{table}[h!]
\centering
\caption{\label{tab:qem_observable_comparison}Averaged staggered magnetization of ten repetition with time when $N = 20$ on \texttt{ibm\_yonsei} under NOQEM and TREX+DD+PT conditions and OBC and PBC. The $\pm$ terms represent the standard deviation of the data. 
}

\vspace{2mm}
\begin{tabular}{clcccc}
\toprule
& & \multicolumn{2}{c}{\textbf{OBC}} & \multicolumn{2}{c}{\textbf{PBC}} \\
\cmidrule(lr){3-4} \cmidrule(lr){5-6}
\textbf{\makecell{Trotter\\Steps}} & & 
\makecell{NOQEM} & TREX+DD+PT & 
\makecell{NOQEM} & TREX+DD+PT \\
\midrule
1 & & $-0.2387 \pm 0.0030$ & $-0.3172 \pm 0.0101$ & $-0.2450 \pm 0.0012$ & $-0.3165 \pm 0.0275$ \\
2 & & $-0.0286 \pm 0.0009$ & $-0.1107 \pm 0.0124$ & $0.0086 \pm 0.0016$ & $-0.0836 \pm 0.0332$ \\
3 & & $0.0710 \pm 0.0010$ & $0.0452 \pm 0.0067$ & $0.0915 \pm 0.0009$ & $0.0588 \pm 0.0149$ \\
4 & & $0.0121 \pm 0.0016$ & $0.0877 \pm 0.0029$ & $0.0166 \pm 0.0021$ & $0.0873 \pm 0.0237$ \\
5 & & $-0.0416 \pm 0.0025$ & $0.0525 \pm 0.0076$ & $-0.0582 \pm 0.0020$ & $0.0390 \pm 0.0316$ \\
6 & & $-0.0363 \pm 0.0026$ & $-0.0064 \pm 0.0047$ & $-0.0504 \pm 0.0008$ & $-0.0205 \pm 0.0109$ \\
7 & & $-0.0114 \pm 0.0018$ & $-0.0349 \pm 0.0067$ & $-0.0060 \pm 0.0013$ & $-0.0394 \pm 0.0115$ \\
8 & & $-0.0039 \pm 0.0014$ & $-0.0264 \pm 0.0043$ & $0.0138 \pm 0.0012$ & $-0.0253 \pm 0.0131$ \\
9 & & $-0.0102 \pm 0.0006$ & $-0.0032 \pm 0.0025$ & $0.0046 \pm 0.0009$ & $0.0008 \pm 0.0033$ \\
10 & & $-0.0121 \pm 0.0004$ & $0.0112 \pm 0.0019$ & $-0.0066 \pm 0.0010$ & $0.0145 \pm 0.0074$ \\
\bottomrule
\end{tabular}
\end{table}

\begin{table}[h!]
\centering
\caption{\label{tab:circuit_stat}Circuit depth and the number of CNOT gates}
\vspace{2mm}
\begin{tabular}{clcccccccc}
\toprule
& & \multicolumn{4}{c}{\textbf{OBC}} & \multicolumn{4}{c}{\textbf{PBC}} \\
\cmidrule(lr){3-6} \cmidrule(lr){7-10}
\textbf{\makecell{Trotter\\Steps}} & & 
\multicolumn{2}{c}{$N = 20$} & \multicolumn{2}{c}{$N = 104$} &
\multicolumn{2}{c}{$N = 20$} & \multicolumn{2}{c}{$N = 84$} \\
\cmidrule(lr){3-4} \cmidrule(lr){5-6} \cmidrule(lr){7-8} \cmidrule(lr){9-10}
& & \makecell{Circuit\\Depth} & CNOTs & \makecell{Circuit\\Depth} & CNOTs &
\makecell{Circuit\\Depth} & CNOTs & \makecell{Circuit\\Depth} & CNOTs \\
\midrule
1 & & 42 & 87 & 42 & 465 & 42 & 90 & 42 & 378 \\
2 & & 64 & 144 & 64 & 774 & 64 & 150 & 64 & 630 \\
3 & & 86 & 201 & 86 & 1083 & 86 & 210 & 86 & 882 \\
4 & & 108 & 258 & 108 & 1392 & 108 & 270 & 108 & 1134 \\
5 & & 130 & 315 & 130 & 1701 & 130 & 330 & 130 & 1386 \\
6 & & 152 & 372 & 152 & 2010 & 152 & 390 & 152 & 1638 \\
7 & & 174 & 429 & 174 & 2319 & 174 & 450 & 174 & 1890 \\
8 & & 196 & 486 & 196 & 2628 & 196 & 510 & 196 & 2142 \\
9 & & 218 & 543 & 218 & 2937 & 218 & 570 & 218 & 2394 \\
10   & & 240 & 600 & 240 & 3246 & 240 & 630 & 240 & 2646 \\
\bottomrule
\end{tabular}
\end{table}

\begin{table}[h!]
\centering
\caption{\label{tab:qem_mae_conversion_extended}The mean absolute errors of staggered magnetization under various QEM strategies on \texttt{ibm\_yonsei}. The results are shown for OBC with N=20, N = 104 and PBC with N=20, N = 84}

\vspace{2mm}
\begin{tabular}{clcccc}
\toprule
& & \multicolumn{2}{c}{\textbf{OBC}} & \multicolumn{2}{c}{\textbf{PBC}} \\
\cmidrule(lr){3-4} \cmidrule(lr){5-6}
\textbf{\makecell{QEM\\Method}} & & 
\makecell{$N = 20$} & \makecell{$N = 104$} & 
\makecell{$N = 20$} & \makecell{$N = 84$} \\
\midrule
NOQEM              & & $0.08419$ &             & $0.08550$ &              \\
TREX                & & $0.08484$ &             & $0.08896$ &              \\
TREX+DD             & & $0.08248$ &             & $0.08267$ &              \\
TREX+PT             & & $0.05664$ &             & $0.05002$ &              \\
TREX+DD+PT          & & $0.04911$ & $0.05641$   & $0.04962$ & $0.05756$     \\
TREX+DD+PT+ZNE      & & $0.04106$ & $0.04917$   & $0.04123$ & $0.05044$     \\
TREX+DD+PT+SM       & & $0.02760$ & $0.02556$   & $0.02218$ & $0.02832$     \\
\bottomrule
\end{tabular}
\end{table}

\begin{table}[h!]
\centering
    \caption{\label{tab:qem_observable_comparison_largescale}Averaged staggered magnetization of ten repetitions with time when on \texttt{ibm\_yonsei} under ZNE and SM conditions and OBC (N = 104) and PBC (N = 84). The $\pm$ terms represent the standard deviation of the data.}

\vspace{2mm}
\begin{tabular}{clcccc}
\toprule
& & \multicolumn{2}{c}{\textbf{OBC, N = 104}} & \multicolumn{2}{c}{\textbf{PBC, N = 84}} \\
\cmidrule(lr){3-4} \cmidrule(lr){5-6}
\textbf{\makecell{Trotter\\Steps}} & & 
\makecell{ZNE} & SM & 
\makecell{ZNE} & SM \\
\midrule
1  & & $-0.3237 \pm 0.0068$ & $-0.3237 \pm 0.0008$ & $-0.3177 \pm 0.0065$ & $-0.3127 \pm 0.0014$ \\
2  & & $-0.1017 \pm 0.0057$ & $-0.1128 \pm 0.0006$ & $-0.1011 \pm 0.0060$ & $-0.1072 \pm 0.0010$ \\
3  & & $0.0506 \pm 0.0033$ & $0.0644 \pm 0.0007$ & $0.0476 \pm 0.0016$ & $0.0578 \pm 0.0008$ \\
4  & & $0.0891 \pm 0.0021$ & $0.1388 \pm 0.0030$ & $0.0862 \pm 0.0020$ & $0.1295 \pm 0.0032$ \\
5  & & $0.0478 \pm 0.0026$ & $0.0878 \pm 0.0016$ & $0.0454 \pm 0.0023$ & $0.0810 \pm 0.0022$ \\
6  & & $-0.0081 \pm 0.0023$ & $-0.0190 \pm 0.0004$ & $-0.0085 \pm 0.0031$ & $-0.0198 \pm 0.0008$ \\
7  & & $-0.0344 \pm 0.0015$ & $-0.0958 \pm 0.0019$ & $-0.0332 \pm 0.0020$ & $-0.0901 \pm 0.0030$ \\
8  & & $-0.0267 \pm 0.0019$ & $-0.0931 \pm 0.0024$ & $-0.0266 \pm 0.0010$ & $-0.0894 \pm 0.0034$ \\
9  & & $-0.0039 \pm 0.0014$ & $-0.0154 \pm 0.0004$ & $-0.0047 \pm 0.0010$ & $-0.0182 \pm 0.0008$ \\
10 & & $0.0121 \pm 0.0013$ & $0.0688 \pm 0.0024$ & $0.0118 \pm 0.0012$ & $0.0674 \pm 0.0037$ \\
\bottomrule
\end{tabular}
\end{table}

\bibliographystyle{iopart-num}
\bibliography{references}

\providecommand{\noopsort}[1]{}\providecommand{\singleletter}[1]{#1}%
\providecommand{\newblock}{}
\begin{thebibliography}{10}
\expandafter\ifx\csname url\endcsname\relax
  \def\url#1{{\tt #1}}\fi
\expandafter\ifx\csname urlprefix\endcsname\relax\def\urlprefix{URL }\fi
\providecommand{\eprint}[2][]{\url{#2}}

\bibitem{google2023suppressing}
 2023 {\em Nature\/} {\bf 614} 676--681

\bibitem{acharya2024quantum}
Acharya R, Abanin D~A, Aghababaie-Beni L, Aleiner I, Andersen T~I, Ansmann M, Arute F, Arya K, Asfaw A, Astrakhantsev N {\em et~al.\/} 2024 {\em Nature\/}

\bibitem{putterman2025hardware}
Putterman H, Noh K, Hann C~T, MacCabe G~S, Aghaeimeibodi S, Patel R~N, Lee M, Jones W~M, Moradinejad H, Rodriguez R {\em et~al.\/} 2025 {\em Nature\/} {\bf 638} 927--934

\bibitem{endo-error-mitigation}
{Endo} S, {Benjamin} S~C and {Li} Y 2018 {\em Physical Review X\/} {\bf 8} 031027 (\textit{Preprint} \eprint{1712.09271})

\bibitem{Temme-error-mitigation}
{Temme} K, {Bravyi} S and {Gambetta} J~M 2017 {\em Physcial Review Letters\/} {\bf 119} 180509 (\textit{Preprint} \eprint{1612.02058})

\bibitem{Li-error-mitigation}
{Li} Y and {Benjamin} S~C 2016 {\em arXiv e-prints\/} arXiv:1611.09301 (\textit{Preprint} \eprint{1611.09301})

\bibitem{Kandala-error-mitigation}
{Kandala} A, {Temme} K, {C{\'o}rcoles} A~D, {Mezzacapo} A, {Chow} J~M and {Gambetta} J~M 2019 {\em Nature\/} {\bf 567} 491--495 (\textit{Preprint} \eprint{1805.04492})

\bibitem{Berg-error-mitigation}
{van den Berg} E, {Minev} Z~K, {Kandala} A and {Temme} K 2023 {\em Nature Physics\/} {\bf 19} 1116--1121 (\textit{Preprint} \eprint{2201.09866})

\bibitem{yu2023simulating}
{Yu} H, {Zhao} Y and {Wei} T~C 2023 {\em Physical Review Research\/} {\bf 5} 013183 (\textit{Preprint} \eprint{2207.09994})

\bibitem{Kim-error-mitigation}
Kim Y, Wood C~J, Yoder T~J, Merkel S~T, Gambetta J~M, Temme K and Kandala A 2023 {\em Nature Phys.\/} {\bf 19} 752--759 (\textit{Preprint} \eprint{2108.09197})

\bibitem{kim2023evidence}
{Kim} Y, {Eddins} A, {Anand} S, {Wei} K~X, {van den Berg} E, {Rosenblatt} S, {Nayfeh} H, {Wu} Y, {Zaletel} M, {Temme} K and {Kandala} A 2023 {\em Nature\/} {\bf 618} 500--505

\bibitem{rahman2022self}
Rahman S~A, Lewis R, Mendicelli E and Powell S 2022 {\em Physical Review D\/} {\bf 106} 074502 (\textit{Preprint} \eprint{2205.09247})

\bibitem{giurgica2020digital}
Giurgica-Tiron T, Hindy Y, LaRose R, Mari A and Zeng W~J 2020 {\em 2020 IEEE International Conference on Quantum Computing and Engineering (QCE)\/}  306--316 (\textit{Preprint} \eprint{2005.10921})

\bibitem{chowdhury2024enhancing}
Chowdhury T~A, Yu K, Shamim M~A, Kabir M and Sufian R~S 2024 {\em Physical Review Research\/} {\bf 6} 033107

\bibitem{Orbach}
Orbach R 1958 {\em Physical Review\/} {\bf 112}(2) 309--316 \urlprefix\url{https://link.aps.org/doi/10.1103/PhysRev.112.309}

\bibitem{Lieb-Schultz-Mattis}
Lieb E, Schultz T and Mattis D 1961 {\em Annals of Physics\/} {\bf 16} 407--466 ISSN 0003-4916 \urlprefix\url{https://www.sciencedirect.com/science/article/pii/0003491661901154}

\bibitem{Gaudin}
Des~Cloizeaux J and Gaudin M 1966 {\em Journal of Mathematical Physics\/} {\bf 7} 1384--1400 ISSN 0022-2488 \urlprefix\url{https://doi.org/10.1063/1.1705048}

\bibitem{Yang-Yang}
Yang C~N and Yang C~P 1966 {\em Physical Review\/} {\bf 150}(1) 321--327 \urlprefix\url{https://link.aps.org/doi/10.1103/PhysRev.150.321}

\bibitem{Alcaraz}
Alcaraz F~C, Barber M~N and Batchelor M~T 1987 {\em Phys. Rev. Lett.\/} {\bf 58}(8) 771--774 \urlprefix\url{https://link.aps.org/doi/10.1103/PhysRevLett.58.771}

\bibitem{DEVEGA1985439}
{de Vega} H and Woynarovich F 1985 {\em Nuclear Physics B\/} {\bf 251} 439--456 ISSN 0550-3213 \urlprefix\url{https://www.sciencedirect.com/science/article/pii/0550321385902718}

\bibitem{Luther-Peschel}
Luther A and Peschel I 1975 {\em Phys. Rev. B\/} {\bf 12}(9) 3908--3917 \urlprefix\url{https://link.aps.org/doi/10.1103/PhysRevB.12.3908}

\bibitem{LUKYANOV1998533}
Lukyanov S 1998 {\em Nuclear Physics B\/} {\bf 522} 533--549 ISSN 0550-3213 \urlprefix\url{https://www.sciencedirect.com/science/article/pii/S0550321398002491}

\bibitem{Calabrese-Cardy-1}
{Calabrese} P and {Cardy} J 2005 {\em Journal of Statistical Mechanics: Theory and Experiment\/} {\bf 2005} 04010 (\textit{Preprint} \eprint{cond-mat/0503393})

\bibitem{DeChiara}
{DeChiara} G, {Montangero} S, {Calabrese} P and {Fazio} R 2006 {\em Journal of Statistical Mechanics: Theory and Experiment\/} {\bf 2006} 03001 (\textit{Preprint} \eprint{cond-mat/0512586})

\bibitem{Fagotti}
{Fagotti} M, {Collura} M, {Essler} F~H~L and {Calabrese} P 2014 {\em Physical Review B\/} {\bf 89} 125101 (\textit{Preprint} \eprint{1311.5216})

\bibitem{Bonnes}
{Bonnes} L, {Essler} F~H~L and {L{\"a}uchli} A~M 2014 {\em Physical Review Letters\/} {\bf 113} 187203 (\textit{Preprint} \eprint{1404.4062})

\bibitem{Essler-review}
{Essler} F~H~L and {Fagotti} M 2016 {\em Journal of Statistical Mechanics: Theory and Experiment\/} {\bf 6} 064002 (\textit{Preprint} \eprint{1603.06452})

\bibitem{trotter}
Trotter H~F 1959 {\em Proceedings of the American Mathematical Society\/} {\bf 10} 545--551 ISSN 00029939, 10886826 \urlprefix\url{http://www.jstor.org/stable/2033649}

\bibitem{suzuki1}
Suzuki M 1976 {\em Communications in Mathematical Physics\/} {\bf 51} 183--190

\bibitem{suzuki2}
Suzuki M 1977 {\em Communications in Mathematical Physics\/} {\bf 57} 193--200

\bibitem{vanicat2018integrable}
Vanicat M, Zadnik L and Prosen T 2018 {\em Phys. Rev. Lett.\/} {\bf 121} 030606

\bibitem{smith2019simulating}
Smith A, Kim M~S, Pollmann F and Knolle J 2019 {\em npj Quantum Inf.\/} {\bf 5} 106 (\textit{Preprint} \eprint{1906.06343})

\bibitem{zhang2024optimal}
Zhang K, Yu K, Hao K and Korepin V 2024 {\em Advanced Quantum Technologies\/}  2300345

\bibitem{kivlichan2020improved}
{Kivlichan} I~D, {Gidney} C, {Berry} D~W, {Wiebe} N, {McClean} J, {Sun} W, {Jiang} Z, {Rubin} N, {Fowler} A, {Aspuru-Guzik} A, {Neven} H and {Babbush} R 2020 {\em Quantum\/} {\bf 4} 296 (\textit{Preprint} \eprint{1902.10673})

\bibitem{lee2021even}
{Lee} J, {Berry} D~W, {Gidney} C, {Huggins} W~J, {McClean} J~R, {Wiebe} N and {Babbush} R 2021 {\em PRX Quantum\/} {\bf 2} 030305 (\textit{Preprint} \eprint{2011.03494})

\bibitem{charles2305simulating}
{Charles} C, {Gustafson} E~J, {Hardt} E, {Herren} F, {Hogan} N, {Lamm} H, {Starecheski} S, {Van de Water} R~S and {Wagman} M~L {Simulating $\mathbb{Z}_2$ lattice gauge theory on a quantum computer} (\textit{Preprint} \eprint{2305.02361})

\bibitem{PhysRevA.105.032620}
van~den Berg E, Minev Z~K and Temme K 2022 {\em Phys. Rev. A\/} {\bf 105}(3) 032620 \urlprefix\url{https://link.aps.org/doi/10.1103/PhysRevA.105.032620}

\bibitem{viola1999dynamical}
Viola L, Knill E and Lloyd S 1999 {\em Phys. Rev. Lett.\/} {\bf 82} 2417--2421 (\textit{Preprint} \eprint{quant-ph/9809071})

\bibitem{ezzell2022dynamical}
Ezzell N, Pokharel B, Tewala L, Quiroz G and Lidar D~A 2023 {\em Phys. Rev. Applied\/} {\bf 20} 064027 (\textit{Preprint} \eprint{2207.03670})

\bibitem{niu2022effects}
Niu S and Todri-Sanial A 2022 {\em IEEE Trans. Quantum Eng.\/} {\bf 3} 1--10 (\textit{Preprint} \eprint{2204.01471})

\bibitem{bennett1996purification}
{Bennett} C~H, {Brassard} G, {Popescu} S, {Schumacher} B, {Smolin} J~A and {Wootters} W~K 1996 {\em Physcial Review Letters\/} {\bf 76} 722--725 (\textit{Preprint} \eprint{quant-ph/9511027})

\bibitem{wallman2016noise}
{Wallman} J~J and {Emerson} J 2016 {\em Physcial Review A\/} {\bf 94} 052325 (\textit{Preprint} \eprint{1512.01098})

\bibitem{cai2019constructing}
Cai Z and Benjamin S~C 2019 {\em Scientific Reports\/} {\bf 9} ISSN 2045-2322 (\textit{Preprint} \eprint{1807.04973}) \urlprefix\url{http://dx.doi.org/10.1038/s41598-019-46722-7}

\bibitem{chowdhury2024capturing}
Chowdhury T~A, Yu K, Asaduzzaman M and Sufian R~S 2024 {\em arXiv preprint arXiv:2412.15180\/}

\bibitem{chowdhury2025first}
Chowdhury T~A, Yu K and Sufian R~S 2025 {\em arXiv preprint arXiv:2503.18580\/}

\bibitem{urbanek2021mitigating}
Urbanek M, Nachman B, Pascuzzi V~R, He A, Bauer C~W and de~Jong W~A 2021 {\em Physical review letters\/} {\bf 127} 270502

\bibitem{quspin}
Weinberg P and Bukov M 2017 {\em SciPost Phys.\/} {\bf 2} 003 \urlprefix\url{https://scipost.org/10.21468/SciPostPhys.2.1.003}

\bibitem{Paeckel:2019yjf}
Paeckel S, K\"ohler T, Swoboda A, Manmana S~R, Schollw\"ock U and Hubig C 2019 {\em Annals Phys.\/} {\bf 411} 167998 (\textit{Preprint} \eprint{1901.05824})

\bibitem{Cirac:2020obd}
Cirac J~I, Perez-Garcia D, Schuch N and Verstraete F 2021 {\em Rev. Mod. Phys.\/} {\bf 93} 045003 (\textit{Preprint} \eprint{2011.12127})

\bibitem{TDVP-ref-1}
{Haegeman} J, {Cirac} J~I, {Osborne} T~J, {Pi{\v{z}}orn} I, {Verschelde} H and {Verstraete} F 2011 {\em Physcial Review Letters\/} {\bf 107} 070601 (\textit{Preprint} \eprint{1103.0936})

\bibitem{TDVP-ref-2}
Haegeman J, Lubich C, Oseledets I, Vandereycken B and Verstraete F 2016 {\em Physical Review B\/} {\bf 94} ISSN 2469-9969 \urlprefix\url{http://dx.doi.org/10.1103/PhysRevB.94.165116}

\bibitem{itensor-1}
Fishman M, White S~R and Stoudenmire E~M 2022 {\em SciPost Phys. Codebases\/}  4 \urlprefix\url{https://scipost.org/10.21468/SciPostPhysCodeb.4}

\bibitem{itensor-2}
Fishman M, White S~R and Stoudenmire E~M 2022 {\em SciPost Phys. Codebases\/}  4--r0.3 \urlprefix\url{https://scipost.org/10.21468/SciPostPhysCodeb.4-r0.3}

\bibitem{itensor-TDVP}
{https://github.com/ITensor/ITensorTDVP.jl.git} \urlprefix\url{https://github.com/ITensor/ITensorTDVP.jl.git}

\bibitem{childs2021theory}
Childs A~M, Su Y, Tran M~C, Wiebe N and Zhu S 2021 {\em Physical Review X\/} {\bf 11} 011020

\bibitem{coote2025resource}
Coote P, Dimov R, Maity S, Hartnett G~S, Biercuk M~J and Baum Y 2025 {\em PRX Quantum\/} {\bf 6} 010332

\bibitem{tong2024empirical}
Tong C, Zhang H and Pokharel B 2024 {\em arXiv preprint arXiv:2403.02294\/}

\bibitem{ezzell2023dynamical}
Ezzell N, Pokharel B, Tewala L, Quiroz G and Lidar D~A 2023 {\em Physical Review Applied\/} {\bf 20} 064027

\bibitem{liu2013noise}
Liu G~Q, Po H~C, Du J, Liu R~B and Pan X~Y 2013 {\em Nature communications\/} {\bf 4} 2254

\bibitem{Buhrman:2001rma}
Buhrman H, Cleve R, Watrous J and de~Wolf R 2001 {\em Phys. Rev. Lett.\/} {\bf 87} 167902 (\textit{Preprint} \eprint{quant-ph/0102001})

\bibitem{Horodecki-swap}
{Horodecki} P and {Ekert} A 2002 {\em Physcial Review Letters\/} {\bf 89} 127902 (\textit{Preprint} \eprint{quant-ph/0111064})

\bibitem{Ekert-swap}
Ekert A~K, Alves C~M, Oi D~K~L, Horodecki M, Horodecki P and Kwek L~C 2002 {\em Phys. Rev. Lett.\/} {\bf 88}(21) 217901 (\textit{Preprint} \eprint{quant-ph/0203016}) \urlprefix\url{https://link.aps.org/doi/10.1103/PhysRevLett.88.217901}

\bibitem{Brydges-randomized}
{Brydges} T, {Elben} A, {Jurcevic} P, {Vermersch} B, {Maier} C, {Lanyon} B~P, {Zoller} P, {Blatt} R and {Roos} C~F 2019 {\em Science\/} {\bf 364} 260--263 (\textit{Preprint} \eprint{1806.05747})

\bibitem{Enk-Beenakker}
van Enk S~J and Beenakker C~W~J 2012 {\em Phys. Rev. Lett.\/} {\bf 108}(11) 110503 (\textit{Preprint} \eprint{1112.1027}) \urlprefix\url{https://link.aps.org/doi/10.1103/PhysRevLett.108.110503}

\bibitem{Elben-1}
{Elben} A, {Vermersch} B, {Dalmonte} M, {Cirac} J~I and {Zoller} P 2018 {\em Physcial Review Letters\/} {\bf 120} 050406 (\textit{Preprint} \eprint{1709.05060})

\bibitem{Vermersch-1}
{Vermersch} B, {Elben} A, {Dalmonte} M, {Cirac} J~I and {Zoller} P 2018 {\em Physcial Review A\/} {\bf 97} 023604 (\textit{Preprint} \eprint{1801.00999})

\bibitem{Elben-2}
{Elben} A, {Vermersch} B, {Roos} C~F and {Zoller} P 2019 {\em Physcial Review A\/} {\bf 99} 052323 (\textit{Preprint} \eprint{1812.02624})

\bibitem{Rath}
{Rath} A, {van Bijnen} R, {Elben} A, {Zoller} P and {Vermersch} B 2021 {\em Physcial Review Letters\/} {\bf 127} 200503 (\textit{Preprint} \eprint{2102.13524})

\bibitem{Elben-3}
{Elben} A, {Flammia} S~T, {Huang} H~Y, {Kueng} R, {Preskill} J, {Vermersch} B and {Zoller} P 2023 {\em Nature Reviews Physics\/} {\bf 5} 9--24 (\textit{Preprint} \eprint{2203.11374})

\bibitem{park2023quantum}
{Park} G, {Zhang} K, {Yu} K and {Korepin} V 2023 {\em Quantum Information Processing\/} {\bf 22} 54 (\textit{Preprint} \eprint{2207.14464})

\bibitem{rao2024quantum}
Rao P, Choi S and Yu K 2024 {\em Proc. SPIE Int. Soc. Opt. Eng.\/} {\bf 12911} 129110E

\bibitem{baker2024parallel}
Baker J~S, Park G, Yu K, Ghukasyan A, Goktas O and Radha S~K 2024 {\em Quantum Machine Intelligence\/} {\bf 6} 18 (\textit{Preprint} \eprint{2305.05881})

\bibitem{huang2020predicting}
Huang H~Y, Kueng R and Preskill J 2020 {\em Nature Physics\/} {\bf 16} 1050--1057

\end{thebibliography}

\end{document}